\documentclass[journal,transaction]{IEEEtran}

\usepackage{float}
\usepackage{tabularx,colortbl}
\usepackage{setspace}
\usepackage{cite,graphicx,amsmath,bm}
\usepackage{subfigure}
\usepackage[usenames,dvipsnames]{xcolor}
\usepackage{mathabx}
\usepackage{cancel}
\usepackage{epstopdf}
\usepackage{amsmath}
\usepackage{xcolor}

\usepackage{bm}

\newcommand{\figref}{Fig.~\ref}

\newcommand{\ve}[1]{\mathbf{#1}}
\newcommand{\uve}[1]{\mathbf{\hat{#1}}}
\newcommand{\ves}[1]{\boldsymbol{#1}}
\newcommand{\te}[1]{\overline{\overline{#1}}}

\definecolor{mygreen}{rgb}{0.4,0.74,0.27}
\definecolor{myorange}{rgb}{0.96,0.47,0.13}
\definecolor{mypurple}{rgb}{0.83,0.39,0.65}
\definecolor{myblue}{rgb}{0.18,0.26,0.61}

\ifCLASSINFOpdf
\else
\fi

\begin{document}

\title{Synthesis  of Spherical Metasurfaces \\ based on Susceptibility Tensor GSTCs }

\author{Xiao~Jia,
        Yousef~Vahabzadeh,
        Fan~Yang,~\IEEEmembership{Senior~Member,~IEEE,}
        Christophe~Caloz,~\IEEEmembership{Fellow,~IEEE}
\thanks{Xiao Jia is with Department of Electronic Engineering, Tsinghua University, Beijing 100084, China and the Department of Electrical Engineering, Polytechnique Montr\'{e}al, Montr\'{e}al, Qu\'{e}bec. (e-mail: jiax16@mails.tsinghua.edu.cn).}
\thanks{Y. Vahabzadeh and C. Caloz are with the Department of Electrical Engineering, Polytechnique Montr\'{e}al,
Montr\'{e}al, Qu\'{e}bec. (e-mail:christophe.caloz@polymtl.ca; yousef.vahabzadeh@polymtl.ca).}
\thanks{Fan Yang is with Department of Electronic Engineering, Tsinghua University, Beijing 100084, China. (e-mail:fan\_yang@tsinghua.edu.cn).}
 }

\maketitle

\begin{abstract}
The bianisotropic susceptibility Generalized Sheet Transition Conditions (GSTCs) synthesis method is extended from planar to spherical metasurfaces. Properties specific to the non-zero intrinsic curvature of the spherical shape are highlighted and different types of corresponding transformations are described. Finally, the susceptibility-GSTC method and exotic properties of spherical metasurfaces are validated and illustrated with three examples: illusion transformation, ring focusing and birefringence.
\end{abstract}


\IEEEdisplaynontitleabstractindextext

\IEEEpeerreviewmaketitle


\section{Introduction}\label{sec:intro}

Metasurfaces are electrically thin 2D structures consisting of a subwavelength lattices of scattering particles that are capable to transform electromagnetic waves in unprecedented fashions~\cite{book_cui2010metamaterials,light_principle_yu2011,overview_metasurface_Holloway_2012,synthesis_planar_KA2015,Tretyakov2016metasurfaces}. Despite their recent emergence, they have already lead to an impressive number of applications, including ultra-thin optical lenses~\cite{lin2014gradient_optic}, high-resolution holograms~\cite{Holography2012}, enhanced classical/quantum efficiency cavities~\cite{chen_Luzhou2017simultaneous}, spatial angular filters~\cite{spatial_fieltering2013}, perfect absorbers~\cite{Single_layer_absorption}, remote controllers~\cite{KA_remote_control2016metasurface}, spatial operators~\cite{computing2014metamaterial}, ultrafast processors~\cite{dataprocess2009ultrafast} and surface plasmonic sensors~\cite{sensors2005generalized}.

The vast majority of metasurfaces reported to date were \emph{planar}. However, many applications, such as cloaking, aircraft RCS reduction, vital signal detection, etc. would greatly benefit form \emph{other metasurface shapes}. Shapes of interest may greatly vary, and even include most complex irregular shapes. However, irregular shapes typically do not admit mathematical solutions and may often be approximated by simpler and more insightful \emph{canonical} shapes. Therefore, it makes sense to first consider such canonical shapes. Canonical shapes may be classified into two main categories~\cite{Ulf_light_and_geo2012}: a)~shapes of zero intrinsic curvature, that may be obtained by folding flat sheets, such as corrugated surfaces, and cylinders with circular, elliptic, parabolic or hyperbolic cross sections, and b)~shapes of non-zero-intrinsic curvature, which are fundamentally 3D, such as spheroidal, prolate, oblate, paraboloidal and conical shells.

The simplest and most common of these canonical shapes are the cylinder with circular cross section, within category a), and the sphere, within category b). The former has been reported in leaky-wave antennas~\cite{cylinder_leakywave_balanis_2017}, 2D beam formers~\cite{cylinder_pattern_control_BR_2017} and RCS reducers~\cite{cylinder_RCS_ruduction}, while the later has been reported in 3D beam formers~\cite{sphere_patterns_impedance_Raeker2016} and antenna decouplers~\cite{sphere_surface_impedance_2014}. The development of practical cylindrical and spherical metasurfaces will naturally require efficient design methods, and the \emph{Bianisotropic Susceptibility - Generalized Sheet Transition Condition} (GSTC) approach~\cite{Averaged_GSTCs_Kuester2003,GSTCs_Idemen2011,synthesis_planar_KA2015}, that has been successfully applied to planar metasurfaces~\cite{synthesis_planar_KA2015}, is a natural candidate for the design of such metasurfaces. The cylindrical case has already been treated in~\cite{cylinder_susceptibility_Capolino2017}. This paper presents the GSTC synthesis of \emph{spherical} metasurfaces.

The paper is organized as follows. Section~\ref{sec:problem} describes the spherical metasurface problem and its specificities. Next, Sec.~\ref{sec:synthesis} presents the extension of the susceptibility-GSTC method to spherical metasurfaces, with specific transformation types and scattering parameter mapping. Then, Sec.~\ref{sec:Ill_ex} illustrates and validates the method for some interesting spherical metasurface transformations. Conclusions are given in Sec.~\ref{sec:concl}.

\section{Spherical Metasurface Problem}\label{sec:problem}

\begin{figure}[!h]
    \centering
         \includegraphics[width=0.7\linewidth, trim={-0.1in 0in -0.4in 0in}]{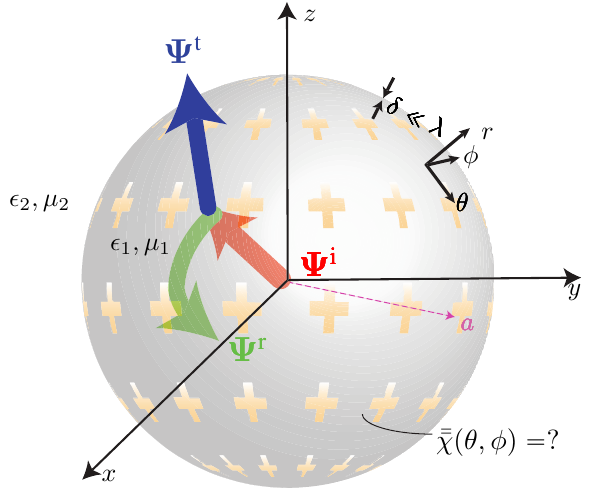}{
        }
        \caption{Spherical metasurface structure and synthesis problem. The structure consists of a deeply subwavelength ($\delta\ll\lambda$) spherical shell of radius $a$, centered at $r=0$, and composed of spherically curved subwavelength scattering particles. The synthesis problem consists in determining the metasurface susceptibility tensor, $\te{\chi}(\theta,\phi)$, for transforming a specified arbitrary incident field, $\ve{\Psi}^\text{i}$, into a specified arbitrary reflected field, $\ve{\Psi}^\text{r}$, and a specified arbitrary transmitted field, $\ve{\Psi}^\text{t}$.}
   \label{FIG:sphere}
\end{figure}

The spherical metasurface structure and synthesis problem are represented in Fig.~\ref{FIG:sphere}. A spherical metasurface is fundamentally different from a planar metasurface, or from a curved metasurface with electrically large curvature\footnote{Under this condition, the Rayleigh hypothesis, according to which scattering is exclusively composed of \emph{outgoing} waves~\cite{Rayleigh_hypothesis}~\cite{rayleigh1896theory}, holds. This practically means that a ray impinging on the metasurface directly scatters (reflects, refracts and/or diffracts) at its incidence point and does not get trapped and multiply scattered in the troughs of the structure. An example of such a curved metasurface is a periodic corrugated metasurface with a ratio of corrugation height over period much smaller than the wavelength.\label{footnote:Rayleigh}}, which may be considered as electromagnetically quasi-planar. In the planar and quasi-planar cases, an incident wave is simply reflected and transmitted by the metasurface in a single scattering event, since the metasurface structure is open and smoothly varying. In contrast, a spherical metasurface is a \emph{closed structure}, forming de facto a \emph{porous cavity}, where the initial reflection-transmission event may be followed by multiple scattering events, and even resonance effects in the case of large local reflections\footnote{That would for instance be the case in the stop-bands of a spherical frequency selective surface (FSS), where that cavity becomes completely opaque.}. Another fundamental difference, is that, while the planar and curved metasurfaces are practically finite and may hence diffract the incident wave at their edges and corners, the spherical metasurface is rotationally infinite and hence does not include edge or corner diffraction\footnote{This is true in the \emph{full} spherical metasurface considered in this paper. However, a \emph{spherical metasurface cap}, which would represent a further interesting and practically useful problem, would naturally also feature edge and corner diffraction.}.

The problem of a \emph{general} porous cavity is very complex, and we do not address it here\footnote{This problem, and other related interesting problems, such as that of a spherical-shell cavity, would naturally require more intensive studies, which are out of the scope of this paper and may be the object of later publications.}. Here, we restrict our attention to the particular case of a spherical metasurface that a)~is excited from its inside, and b)~is reflection-less, so as to avoid multiple internal scattering\footnote{This problem may be considered as the limiting case of a porous spherical cavities with full porosity.} as planar and quasi-planar metasurfaces. While this is a major restriction, the inside-excitation reflection-less problem already represents an electromagnetically rich and practically interesting metasurface, allowing unusual and exotic field transformations, as will be shown next.

\begin{figure}[!h]
    \centering
         \includegraphics[width=1\linewidth, trim={-0.1in 0in -0.1in 0in}]{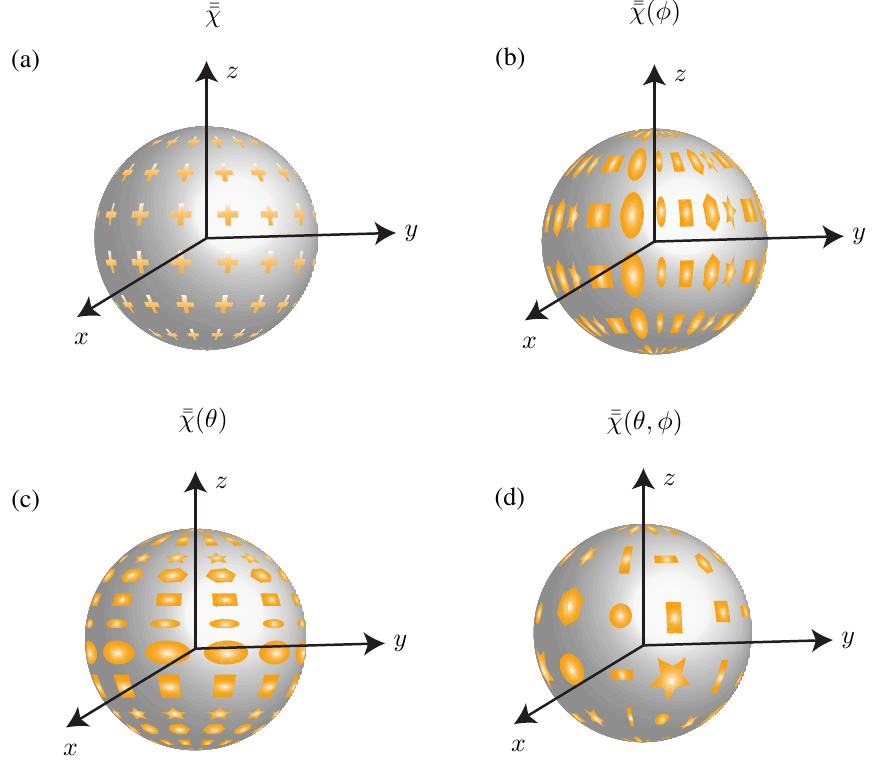}{
        }
        \caption{Categories of spherical metasurfaces based on their susceptibilities. (a)~Uniform susceptibility. (b)~$\theta$-uniform and $\phi$-varying susceptibitity. (c)~$\phi$-uniform and $\theta$-varying susceptibitity (same problem as b), upon $\pi/2$ rotation). (d)~Double nonuniform susceptibitity.}
   \label{FIG:class}
\end{figure}

Figure~\ref{FIG:class} shows the different possible categories of spherical metasurfaces based on their azimuthal and elevation susceptibility variations. The susceptibility may be uniform in both elevation and azimuth [Fig.~\ref{FIG:class}(a)], uniform in one angular direction and nonuniform in the other one [Figs.~\ref{FIG:class}(b) and~(c)], or nonuniform in both angular directions [Fig.~\ref{FIG:class}(d)].


\section{Synthesis}\label{sec:synthesis}
\subsection{Susceptibility GSTCs}\label{sec:synthesis_gstc}

The GSTC equations for the spherical metasurface problem in~\figref{FIG:sphere} are derived in Appendix~\ref{GSTC_Deriv}. They may be written as
\begin{subequations}\label{Eq:BC}
\begin{align}
[\uve{r}\times\triangle\ve{H}&=\ve{J}_\text{s,tot}]_{r=a},\\
[\uve{r}\times\triangle\ve{E}&=-\ve{K}_\text{s,tot}]_{r=a},
\end{align}
\end{subequations}
where $\uve{r}$ is the unit vector normal the metasurface.
In these relations, the $\triangle$ symbol represents the field jumps at the metasurface discontinuity, i.e.
\begin{equation}\label{Eq:diff_fields}
\triangle\ves{\Psi}
=\ves{\Psi}^+-\ves{\Psi}^-
=\ves{\Psi}^\text{t}-(\ves{\Psi}^\text{i}+\ves{\Psi}^\text{r}),
\quad
\ves{\Psi}=\ve{E},\ve{H},
\end{equation}
with the superscripts $\pm$ referring to the points just before and just after the metasurface, i.e. at $r=a^\pm$, and the superscripts i, r and t referring to the incident, reflected and transmitted fields, respectively. Moreover,
\begin{subequations}\label{Eq:JKtotal}
\begin{align}\label{Eq:Jtotal}
\ve{J}_\text{s,tot}
&=\ve{J}_\text{s,imp}+\ve{J}_\text{s,p}+\ve{J}_\text{s,m}\\
&=\ve{J}_\text{s,imp}+\frac{\partial \ve{P}_\text{s}}{\partial t}+\nabla\times\ve{M}_\text{s}
\quad\text{(A/m)}
\end{align}
and
\begin{align}\label{Eq:Ktotal}
\ve{K}_\text{s,tot}
&=\ve{K}_\text{s,imp}+\ve{K}_\text{s,m}+\ve{K}_\text{s,p}\\
&=\ve{K}_\text{s,imp}+\mu_0 \frac{\partial \ve{M}_\text{s}}{\partial t}+\nabla \times(\ve{P}_\text{s}/\epsilon_0)
\quad\text{(V/m)}
\end{align}
\end{subequations}
represent the total \emph{surface} current densities on the metasurface, which are composed of the impressed electric and magnetic surface current densities, $\ve{J}_\text{s,imp}$ and $\ve{K}_\text{s,imp}$, the electric surface current densities due to electric and magnetic polarization densities, $\ve{J}_\text{s,p}$ and $\ve{J}_\text{s,m}$, and the magnetic surface current densities due to electric and magnetic polarization densities, $\ve{K}_\text{s,p}$ and $\ve{K}_\text{s,m}$. Of course, $\ve{P}_\text{s}$ and $\ve{M}_\text{s}$ in~\eqref{Eq:JKtotal} are \emph{surface} polarization densities (see Appendix~\ref{GSTC_Deriv}), measured in C/m and A, respectively. Note that Eqs.~\eqref{Eq:JKtotal} are restricted to \emph{first-order discontinuities}, with $\ves{\Psi}_\text{s}=\ves{\Psi}\delta(r-a)$, $\ves{\Psi}_\text{s}=\ve{J}_\text{s,p},\ve{K}_\text{s,p},\ve{J}_\text{s,m},\ve{K}_\text{s,m}$\footnote{This restriction is valid for most practical metasurfaces. However, there are cases where it would not be acceptable. For instance, a metasurface transforming the incident field into a transmitted field being to its phase-reversed version could not be described by a series truncated to $N=0$. Indeed, the corresponding fields would include only the even $\delta(r-a)$ distribution whereas the field transformation is obviously odd in nature. In such a case, one should at least include the term $N=1$, to include the odd distribution $\delta'(r-a)$, following~\cite{GSTCs_Idemen2011,synthesis_planar_KA2015}, which would involve extra terms in~\eqref{Eq:JKtotal}. The application of higher-order GSTCs to metasurfaces is still an open research topic.}.

In the particular case $\ve{P}_\text{s}=\ve{M}_\text{s}=0$, only the impressed surface current densities, $\ve{J}_\text{s,imp}$ and $\ve{K}_\text{s,imp}$, survive in~\eqref{Eq:JKtotal}, and Eqs.~\eqref{Eq:BC} reduce to the usual boundary conditions at the interface between two media~\cite{rothwell2008electromagnetics}. However, we are interested here in the opposite case, where $\ve{J}_\text{s,imp}=\ve{K}_\text{s,imp}=0$, assuming the inexistence of sources on the metasurface, and field discontinuities only due to the polarization currents modeling the response of the metasurface scattering particles via the surface electric and magnetic polarization densities $\ve{J}_\text{s,p}$, $\ve{J}_\text{s,m}$, $\ve{K}_\text{s,p}$ and $\ve{K}_\text{s,m}$, respectively\footnote{The GSTCs~\eqref{Eq:BC} may thus be considered as a generalization of the usual boundary conditions including the effect of \emph{surface} material polarization. Equations~\eqref{Eq:JKtotal} were not common in the ``pre-metasurface era'' literature because, before the advent of metasurfaces, 2D materials (e.g. 2DEGs, graphene, etc.) and related computational sheets, polarization was essentially a \emph{volume} concept, defined as the densities of electric and magnetic moments in 3D space, $\ve{P}$ and $\ve{M}$, measured in (C/m$^2$) and (A/m), respectively, which did not make sense in two dimensions.}.

In this paper, we shall restrict our attention to time-harmonic ($e^{j\omega t}$) metasurfaces\footnote{Polychromatic planar time-varying and nonlinear metasurface transformations have been considered in~\cite{Nimaspacetime, KA_nonlinear_2017}.}. Inserting the time-harmonic versions ($\cdot j\omega$) of the polarization current densities~\eqref{Eq:JKtotal}, with $\ve{J}_\text{s,imp}=\ve{K}_\text{s,imp}=0$, into the GSTCs~\eqref{Eq:BC} yields
 \begin{subequations}\label{Eq:GSTC_two_media_JK}
\begin{align}
[\uve{r}\times\triangle\ve{H}&=j\omega\ve{P_{\text{s},\|}}-\uve{r}\times\nabla_{\|}M_{\text{s},r}]_{r=a},\\
[\uve{r}\times\triangle\ve{E}&=-j\omega\mu_0\ve{M_{\text{s},\|}}+\nabla_{\|}(P_{\text{s},r}/\epsilon_0)\times\uve{r}]_{r=a}.
\end{align}
\end{subequations}

The physical metasurface will actually be a spherical \emph{shell} with finite thickness $\delta$, as indicated in~\figref{FIG:sphere}. However, this thickness is typically deeply subwavelength ($\delta\ll\lambda$). Therefore, the shell cannot support significant propagation or resonance effects along the $r$ direction, and the metasurface may hence be safely modelled as a zero-thickness ($\delta=0$) sheet discontinuity through the \emph{bianisotropic surface susceptibility tensor functions} $\te{\chi}_\text{ee}$, $\te{\chi}_\text{mm}$, $\te{\chi}_\text{em}$ and $\te{\chi}_\text{me}$, that relate the average fields at both sides of the metasurface,
\begin{equation}\label{Eq:av_fields}
\ves{\Psi}_\text{av}
=\frac{\Psi^\text{t}+(\Psi^\text{i}+\Psi^\text{r})}{2},
\quad
\ves{\Psi}=\ve{E},\ve{H},
\end{equation}
to the surface polarization densities as~\cite{synthesis_planar_KA2015}
\begin{subequations}\label{Eq:PM}
\begin{align}
 \ve{P}&=\epsilon_0\bar{\bar{\chi}}_\text{ee}\ve{E}_\text{av}+\sqrt{\mu_0\epsilon_0}\bar{\bar{\chi}}_\text{em}\ve{H}_\text{av},\\
 \ve{M}&=\sqrt{\epsilon_0/\mu_0}\bar{\bar{\chi}}_\text{me}\ve{E}_\text{av}+\bar{\bar{\chi}}_\text{mm}\ve{H}_\text{av}.
\end{align}
\end{subequations}
Inserting~\eqref{Eq:PM} into~\eqref{Eq:GSTC_two_media_JK} finally provides the GSTC relations
\begin{subequations}\label{Eq:gstcdeltaav}
\begin{align}
\begin{split}
\uve{r}\times\triangle\ve{H}&=j\omega({\epsilon_0\te{\chi}_\text{ee}\ve{E}_\text{av}+\sqrt{\mu_0\epsilon_0}\te\chi}_\text{em}\ve{H}_\text{av})_\|\\
&-\uve{r}\times\nabla_{\|}[(\sqrt{\epsilon_0/\mu_0}\te{\chi}_\text{me}\ve{E}_\text{av}+\te{\chi}_\text{mm}\ve{H}_\text{av})_r],
\end{split}\\
\begin{split}
\uve{r}\times\triangle\ve{E}&=-j\omega\mu_0(\sqrt{\epsilon_0/\mu_0}\te{\chi}_\text{me}\ve{E}_\text{av}+\te{\chi}_\text{mm}\ve{H}_\text{av})_\|\\
&+\nabla_{\|}[(\epsilon_0\te{\chi}_\text{ee}\ve{E}_\text{av}+\sqrt{\mu_0\epsilon_0}\te{\chi}_\text{em}\ve{H}_\text{av})_r/\epsilon_0]\times\uve{r},
\end{split}
\end{align}
\end{subequations}
explicitly expressed in terms of the difference fields in~\eqref{Eq:diff_fields} on the left-hand sides and average fields in~\eqref{Eq:av_fields} on the right-hand sides through the surface susceptibility tensors, that read in spherical coordinates
\begin{equation}\label{Eq:tensors}
\bar{\bar{\chi}}_\text{ab}(\theta,\phi)=
\begin{bmatrix}
  \chi^{r r}_\text{ab}(\theta,\phi)& \chi^{r \theta}_\text{ab}(\theta,\phi)&\chi^{r \phi}_\text{ab}(\theta,\phi)\\
  \chi^{\theta r}_\text{ab}(\theta,\phi)& \chi^{\theta \theta}_\text{ab}(\theta,\phi)&\chi^{\theta \phi}_\text{ab}(\theta,\phi)\\
  \chi^{\phi r}_\text{ab}(\theta,\phi)& \chi^{\phi \theta}_\text{ab}(\theta,\phi)&\chi^{\phi \phi}_\text{ab}(\theta,\phi)
  \end{bmatrix},
\end{equation}
with $(\text{a,b})=\text{(e,e)}$, $\text{(e,m)}$, $\text{(m,e)}$ and $\text{(m,m)}$. In~\eqref{Eq:gstcdeltaav}, we have dropped, for notational simplicity, the $[\ldots]_{r=a}$ specification, which is implicitly assumed from now on.

This information of the possibly different media surrounding the metasurface is \emph{implicitly} present in~\eqref{Eq:diff_fields} and~\eqref{Eq:av_fields}, as pointed out in Appendix~\ref{GSTC_Deriv}.

\subsection{Synthesis Equations}

In this paper, we shall assume $P_r=M_r=0$, which simplifies the coupled partial differential equations~\eqref{Eq:tensors} to a simple algebraic linear system of equations. As extensively discussed in Secs.~IV.A and~IV.D of~\cite{2017_12_Achouri_Nanophotonics}, this represents a restriction in terms of fabrication and \emph{separate} transformation diversity, but no restriction in terms of an ideal metasurface performing a given transformation, including a transformation involving multiple \emph{simultaneous} operations, since a metasurface with normal polarization components can always be reduced to an equivalent metasurface purely tangential polarization components.

Under the condition $P_r=M_r=0$, Eqs.~\eqref{Eq:gstcdeltaav} with~\eqref{Eq:tensors} reduce to

\begin{subequations}\label{Eq:GSTC_EH}
\begin{equation}\label{Eq:GSTC_expanded_h}
\begin{split}
     \begin{bmatrix}
   -\triangle H_\phi  \\
   \triangle H_\theta
  \end{bmatrix}=j\omega\epsilon_0 &\begin{bmatrix}
 \chi^{\theta\theta}_\text{ee}&\chi^{\theta\phi}_\text{ee}\\
 \chi^{\phi \theta}_\text{ee}&\chi^{\phi \phi}_\text{ee}
  \end{bmatrix}\begin{bmatrix}
   E_{\theta,\text{av}} \\
   E_{\phi,\text{av}}
  \end{bmatrix}\\
  &+j\omega \sqrt{\mu_0 \epsilon_0} \begin{bmatrix}
\chi^{\theta \theta}_\text{em}&\chi^{\theta \phi}_\text{em}\\
 \chi^{\phi \theta}_\text{em}&\chi^{\phi \phi}_\text{em}
  \end{bmatrix} \begin{bmatrix}
                  H_{\theta,\text{av}} \\
                  H_{\phi,\text{av}}
                \end{bmatrix}
\end{split}
\end{equation}
and
\begin{equation}\label{Eq:GSTC_expanded_e}
\begin{split}
     \begin{bmatrix}
   \triangle E_\phi  \\
   -\triangle E_\theta
  \end{bmatrix}=j\omega\mu_0 &\begin{bmatrix}
\chi^{\theta \theta}_\text{mm}&\chi^{\theta \phi}_\text{mm}\\
\chi^{\phi \theta}_\text{mm}&\chi^{\phi \phi}_\text{mm}
  \end{bmatrix}\begin{bmatrix}
   H_{\theta,\text{av}}  \\
   H_{\phi,\text{av}}
  \end{bmatrix}\\
  &+j\omega \sqrt{\mu_0 \epsilon_0} \begin{bmatrix}
\chi^{\theta \theta}_\text{me}&\chi^{\theta\phi}_\text{me}\\
 \chi^{\phi \theta}_\text{me}&\chi^{\phi \phi}_\text{me}
  \end{bmatrix} \begin{bmatrix}
                  E_{\theta,\text{av}} \\
                  E_{\phi,\text{av}}
                \end{bmatrix}.
\end{split}
\end{equation}
\end{subequations}

Equations~\eqref{Eq:GSTC_EH} represent a system of 4 equations in 16 unknowns ($\chi^{\theta \theta}_\text{mm}$, $\chi^{\theta\phi}_\text{ee}$, etc.). So, this is a heavily under-determined system. There are three approaches to solve this problem~\cite{2017_12_Achouri_Nanophotonics}. The first one is to reduce the number of unknowns from 16 to 4, in which case there would be $4^4=256$ possible distinct susceptibility quadruplets, with only a subset of them representing physically meaningful situations. The second approach is to increase the number of simultaneous field transformation specifications from 1 to 4. The last approach is a combination of the first two. To design an optimal metasurface, one has to make an educated choice of approach and susceptibility sets. Such an educated choice includes the conisideration of the following fundamental conditions:
\begin{itemize}
  \item reciprocal (possibly with loss or gain) metasurface:
\begin{equation}\label{reciprocity}
\bar{\bar{\chi}}_\text{ee}^T=\bar{\bar{\chi}}_\text{ee},\quad
\bar{\bar{\chi}}_\text{mm}^T=\bar{\bar{\chi}}_\text{mm},\quad
\bar{\bar{\chi}}_\text{me}^T=-\bar{\bar{\chi}}_\text{em},
   \end{equation}
which implies the suppression of 6 complex (i.e. 12 real numbers) susceptibility degrees of freedom;
\item loss/gain-less reciprocal metasurface:
   \begin{equation}\label{passive_lossless_reciprocal}
\bar{\bar{\chi}}_\text{ee}^T=\bar{\bar{\chi}}_\text{ee}^*,\quad
\bar{\bar{\chi}}_\text{mm}^T=\bar{\bar{\chi}}_\text{mm}^*, \quad
\bar{\bar{\chi}}_\text{me}^T=\bar{\bar{\chi}}_\text{em}^*,
     \end{equation}
which implies the suppression of 16 real number degrees of freedom among the complex susceptibilities;
\item non-gyrotropic metasurface:
 \begin{equation}\label{non-gyrotropy}
\bar{\bar{\chi}}_\text{ee,mm}^{\theta\phi}=0,\;
\bar{\bar{\chi}}_\text{ee,mm}^{\phi\theta}=0,\;
\bar{\bar{\chi}}_\text{em,me}^{\theta\theta}=0,\;
\bar{\bar{\chi}}_\text{em,me}^{\phi\phi}=0,
      \end{equation}
which implies the suppression of 8 complex (i.e. 16 real numbers) susceptibility degrees of freedom;
\end{itemize}

\subsection{Transformation Types}

We derive here closed-form susceptibility solutions to~\eqref{Eq:GSTC_EH} for a few types of transformations depending on the aforementioned approaches. The solutions to other types of transformations can be derived in a similar manner. In this section, we will give these closed-form solutions as functions of the difference fields~\eqref{Eq:diff_fields} and average fields~\eqref{Eq:av_fields}, i.e. as implicit synthesis relations, while explicit examples will be given in Sec.~\ref{sec:Ill_ex}.

\subsubsection{Monoisotropic Transformation}

The simplest possible transformation is the monoisotropic transformation, which may be considered as a particular case of a 4-parameter transformation (approach~1) with $\chi_\text{ee,mm}^{\theta,\theta}=\chi_\text{ee,mm}^{\phi,\phi}=\chi_\text{ee,mm}$. In this case, Eqs.~\eqref{Eq:GSTC_EH} reduce to

\begin{subequations}
\begin{equation}\label{Eq:GSTC_isotropic_e}
     \begin{bmatrix}
   -\triangle H_\phi  \\
   \triangle H_\theta
  \end{bmatrix}=j\omega\epsilon_0 \chi_\text{ee}  \begin{bmatrix}
   E_{\theta,\text{av}} \\
   E_{\phi,\text{av}}
  \end{bmatrix},
\end{equation}

and

\begin{equation}\label{Eq:GSTC_isotropic_h}
     \begin{bmatrix}
   \triangle E_\phi  \\
   -\triangle E_\theta
  \end{bmatrix}=j\omega\mu_0 \chi_\text{mm}    \begin{bmatrix}
   H_{\theta,\text{av}}  \\
   H_{\phi,\text{av}}
  \end{bmatrix}.
\end{equation}
\end{subequations}
Solving this system for $\te{\chi}_\text{ee}$ and $\te{\chi}_\text{mm}$ yields the closed-form synthesis solutions
\begin{subequations}
\begin{align}\label{Eq:GSTC_monoiso_chi}
  \chi_\text{ee} & =-\frac{\triangle H_\phi}{j\omega\epsilon_0 E_{\theta,\text{av}}}=\frac{\triangle H_\theta}{j\omega\epsilon_0 E_{\phi,\text{av}}}, \\
  \chi_\text{mm} & =\frac{\triangle E_\phi}{j\omega\mu_0 H_{\theta,\text{av}}} =-\frac{\triangle E_\theta}{j\omega\mu_0 H_{\phi,\text{av}}},
\end{align}
\end{subequations}
showing that the corresponding metasurface performs the same transformation on $\theta$-polarized and $\phi$-polarized waves.

\subsubsection{Monoanisotropic Transformation}\label{sec:monoanis_transf}

For the metasurface to perform different transformations on the $\theta$- and $\phi$-polarizations (birefringence), one may lift the previous (monoisotropic) restriction to monoanisotropy, involving the susceptibilities $\chi^{\theta \theta}_\text{ee}$, $\chi^{\phi\phi}_\text{ee}$, $\chi^{\theta \theta}_\text{mm}$ and $\chi^{\phi\phi}_\text{mm}$. This is another type of 4-parameter transformation (approach~1), but this time with 4 distinct susceptibilities. In this case, Eqs.~\eqref{Eq:GSTC_EH} become
\begin{subequations}\label{Eq:matrix_mino_ani}
\begin{align}
     &\begin{bmatrix}
   -\triangle H_\phi  \\
   \triangle H_\theta
  \end{bmatrix}=j\omega\epsilon_0 \begin{bmatrix}
 \chi^{\theta\theta}_\text{ee}&0\\
0&\chi^{\phi \phi}_\text{ee}
  \end{bmatrix}\begin{bmatrix}
   E_{\theta,\text{av}} \\
   E_{\phi,\text{av}}
  \end{bmatrix},\\
     &\begin{bmatrix}
   \triangle E_\phi  \\
   -\triangle E_\theta
  \end{bmatrix}=j\omega\mu_0 \begin{bmatrix}
\chi^{\theta \theta}_\text{mm}&0\\
0&\chi^{\phi \phi}_\text{mm}
  \end{bmatrix}\begin{bmatrix}
   H_{\theta,\text{av}}  \\
   H_{\phi,\text{av}}
  \end{bmatrix},
\end{align}
\end{subequations}
and their solution is
\begin{subequations}\label{GSTC_X_e}
  \begin{equation}
  \chi^{\theta \theta}_\text{ee} =-\frac{\triangle H_\phi}{j\omega\epsilon_0 E_{\theta,\text{av}}}, \quad
  \chi^{\phi\phi}_\text{mm} =-\frac{\triangle E_\theta}{j\omega\mu_0 H_{\phi,\text{av}}}.
\end{equation}
  \begin{equation}
  \chi^{\phi\phi}_\text{ee} =\frac{\triangle H_\theta}{j\omega\epsilon_0 E_{\phi,\text{av}}},\quad
  \chi^{\theta \theta}_\text{mm} =\frac{\triangle E_\phi}{j\omega\mu_0 H_{\theta,\text{av}}},
\end{equation}
\end{subequations}
which correspond to $\theta$ and $\phi$ polarizations, respectively.

\subsubsection{Bianisotropic Transformation}

As shown in~\cite{GL_perfect_refraction2017}\cite{perfect_refraction2016}, perfect refraction, i.e. refraction without loss/gain and without spurious diffraction, requires bianisotropy, for which $\te{\chi}_\text{em}\neq 0$ and $\te{\chi}_\text{me}\neq 0$. If one further wishes to perform such a transformation without field rotation (gyrotropy), the condition~\eqref{non-gyrotropy} must be further enforced, which eliminates 8 complex susceptibilities (approach~1). This leaves out 8 complex susceptibility parameters (among which one must ensure $\te{\chi}_\text{em}\neq 0$ and $\te{\chi}_\text{me}\neq 0$), which further reduces to 4~complex susceptibility parameters if one cares for only one polarization (and the transformation of the other polarization is arbitrary). In this case, we also have two equations in~\eqref{Eq:GSTC_EH} disappearing, reducing the total number of equations from 4 to~2. In the case of $\theta$-polarization, we have then 2 equations for the remaining parameters are $\chi^{\theta \theta}_\text{ee}$, $\chi^{\phi \phi}_\text{mm}$, $\chi^{\theta \phi}_\text{em}$ and $\chi^{\phi \theta}_\text{me}$. So, the system is under-determined, which allows us to specify a second transformation (approach~2). In this case, Eqs.~\eqref{Eq:GSTC_EH} may be compactly written
\begin{equation}\label{Eq:bianisotropic}
\begin{small}
     \begin{bmatrix}
\triangle H_{\phi1}&  \triangle H_{\phi2} \\
\triangle E_{\theta1}&  \triangle E_{\theta2}
  \end{bmatrix}= \begin{bmatrix}
 -j\omega\epsilon_0\chi^{\theta\theta}_\text{ee}& -jk_0\chi^{\theta\phi}_\text{em}\\
-jk_0\chi^{\phi\theta}_\text{me}& -j\omega\mu_0\chi^{\phi\phi}_\text{mm}
  \end{bmatrix}
\begin{bmatrix}
 E_{\theta1,\text{av}}&   E_{\theta2,\text{av}} \\
H_{\phi1,\text{av}}&  H_{\phi2,\text{av}}
  \end{bmatrix}
\end{small}
\end{equation}
where the subscripts 1 and 2 correspond to the two transformations. The double transformation in~\eqref{Eq:bianisotropic} involves only 2 of the 4 equations in~\eqref{Eq:GSTC_EH}, and is hence a reduced-rank (from 4 to 2) transformation.

Equation~\eqref{Eq:bianisotropic} represents a system of 4 equations in 4 unknowns, whose solution is
\begin{subequations}\label{Eq:bianisotropicchi}
  \begin{align}
  \chi^{\theta \theta}_\text{ee} & =\frac{1}{j\omega\epsilon_0}\frac{\triangle H_{\phi2} H_{\phi1,\text{av}}-\triangle H_{\phi1} H_{\phi2,\text{av}}}{ E_{\theta1,\text{av}}H_{\phi2,\text{av}}-E_{\theta2,\text{av}}H_{\phi1,\text{av}}},\\
    \chi^{\theta \phi}_\text{em} & =\frac{1}{jk_0}\frac{\triangle H_{\phi2} E_{\theta1,\text{av}}-\triangle H_{\phi1} E_{\theta2,\text{av}}}{ E_{\theta2,\text{av}}H_{\phi1,\text{av}}-E_{\theta1,\text{av}}H_{\phi2,\text{av}}},\\
        \chi^{\phi \theta}_\text{me} & =\frac{1}{jk_0}\frac{\triangle E_{\theta2} H_{\phi1,\text{av}}-\triangle E_{\theta1} H_{\phi2,\text{av}}}{ E_{\theta1,\text{av}}H_{\phi2,\text{av}}-E_{\theta2,\text{av}}H_{\phi1,\text{av}}},\\
            \chi^{\phi \phi}_\text{mm} & =\frac{1}{j\omega\mu_0}\frac{\triangle E_{\theta2} E_{\theta1,\text{av}}-\triangle E_{\theta1} E_{\theta2,\text{av}}}{ E_{\theta2,\text{av}}H_{\phi1,\text{av}}-E_{\theta1,\text{av}}H_{\phi2,\text{av}}}.
\end{align}
\end{subequations}
Equation~\eqref{Eq:bianisotropicchi} generally represents a double transformation. If one further wanted a reciprocal metasurface, as is often the case both functionally and practically, then the third relation in~\eqref{reciprocity} would demand a)~$\chi^{\theta\phi}_\text{em}=-\chi^{\phi\theta}_\text{me}$, and b)~transformation~2 to be the reciprocal transformation of transformation~1\footnote{In this case, transformation~2 would be from the outside to the inside of the metasurface sphere, and the reciprocity specification would be physical only if the wave is strongly attenuated at the metasurface surface or/and within its filling medium.}. The combination of these 2~constraints leads to a new fully-determined system, which may seen as a single reciprocal transformation.

\subsubsection{Full-rank Double Transformation}

The double transformation of~\eqref{Eq:bianisotropic} is a reduced-rank one because it specifies only one polarization specification. We shall now consider the case of a transformation with specifications for \emph{both} polarizations. This leads to a full-rank system, involving the 4 equations in~\eqref{Eq:GSTC_EH} and, without non-gyrotropy constraint, 16 unknowns. Using approach~1, we further specify here monoanisotropy, which leads, using the short-cut notation $\tilde{\chi}_\text{ee}=j\omega\epsilon_0\chi_\text{ee}$ and $\tilde{\chi}_\text{mm}=j\omega\mu_0\chi_\text{mm}$, to the system
\begin{small}
\begin{equation}\label{Eq:doublematrix}
\begin{bmatrix}
-\triangle H_{\phi1}&  -\triangle H_{\phi2} \\
\triangle H_{\theta1}&  \triangle H_{\theta2} \\
\triangle E_{\phi1}&  \triangle E_{\phi2} \\
-\triangle E_{\theta1}&  -\triangle E_{\theta2}
  \end{bmatrix}= \begin{bmatrix}
 \tilde{ \chi}^{\theta\theta}_\text{ee}&\tilde{\chi}^{\theta\phi}_\text{ee}&0&0\\
 \tilde{ \chi}^{\phi \theta}_\text{ee}&\tilde{\chi}^{\phi \phi}_\text{ee}&0&0\\
  0&0&\tilde{\chi}^{\theta\theta}_\text{mm}&\tilde{\chi}^{\theta\phi}_\text{mm}\\
    0&0&\tilde{\chi}^{\phi \theta}_\text{mm}&\tilde{\chi}^{\phi \phi}_\text{mm}
  \end{bmatrix}
\begin{bmatrix}
 E_{\theta1,\text{av}}&   E_{\theta2,\text{av}} \\
 E_{\phi1,\text{av}}&   E_{\phi2,\text{av}} \\
H_{\theta1,\text{av}}&   H_{\theta2,\text{av}} \\
H_{\phi1,\text{av}}&  H_{\phi2,\text{av}}
  \end{bmatrix},
\end{equation}
\end{small}
whose solution is
\begin{subequations}\label{Eq:doublechi}
  \begin{align}
  \chi^{\theta \theta}_\text{ee} & =\frac{1}{j\omega\epsilon_0}\frac{\triangle H_{\phi2} E_{\phi1,\text{av}}-\triangle H_{\phi1} E_{\phi2,\text{av}}}{ E_{\theta1,\text{av}}E_{\phi2,\text{av}}-E_{\theta2,\text{av}}E_{\phi1,\text{av}}},\\
\chi^{\theta \phi}_\text{ee} & =\frac{1}{j\omega\epsilon_0}\frac{\triangle H_{\phi2} E_{\theta1,\text{av}}-\triangle H_{\phi1} E_{\theta2,\text{av}}}{ E_{\phi1,\text{av}}E_{\theta2,\text{av}}-E_{\phi2,\text{av}}E_{\theta1,\text{av}}},\\
  \chi^{\phi \theta}_\text{ee} & =\frac{1}{j\omega\epsilon_0}\frac{\triangle H_{\theta1} E_{\phi2,\text{av}}-\triangle H_{\theta2} E_{\phi1,\text{av}}}{ E_{\theta1,\text{av}}E_{\phi2,\text{av}}-E_{\theta2,\text{av}}E_{\phi1,\text{av}}},\\
\chi^{\phi \phi}_\text{ee} & =\frac{1}{j\omega\epsilon_0}\frac{\triangle H_{\theta1} E_{\theta2,\text{av}}-\triangle H_{\theta2} E_{\theta1,\text{av}}}{ E_{\phi1,\text{av}}E_{\theta2,\text{av}}-E_{\phi2,\text{av}}E_{\theta1,\text{av}}},\\
\chi^{\theta \theta}_\text{mm} & =\frac{1}{j\omega\mu_0}\frac{\triangle E_{\phi1} H_{\phi2,\text{av}}-\triangle E_{\phi2} H_{\phi1,\text{av}}}{ H_{\theta1,\text{av}}H_{\phi2,\text{av}}-H_{\theta2,\text{av}}H_{\phi1,\text{av}}},\\
 \chi^{\theta \phi}_\text{mm} & =\frac{1}{j\omega\mu_0}\frac{\triangle E_{\phi1} H_{\theta2,\text{av}}-\triangle E_{\phi2} H_{\theta1,\text{av}}}{ H_{\phi1,\text{av}}H_{\theta2,\text{av}}-H_{\phi2,\text{av}}H_{\theta1,\text{av}}},\\
  \chi^{\phi \theta}_\text{mm} & =\frac{1}{j\omega\mu_0}\frac{\triangle E_{\theta2} H_{\phi1,\text{av}}-\triangle E_{\theta1} H_{\phi2,\text{av}}}{ H_{\theta1,\text{av}}H_{\phi2,\text{av}}-H_{\theta2,\text{av}}H_{\phi1,\text{av}}},\\
 \chi^{\phi \phi}_\text{mm} & =\frac{1}{j\omega\mu_0}\frac{\triangle E_{\theta2} H_{\theta1,\text{av}}-\triangle E_{\theta1} H_{\theta2,\text{av}}}{ H_{\phi1,\text{av}}H_{\theta2,\text{av}}-H_{\phi2,\text{av}}H_{\theta1,\text{av}}}.
\end{align}
\end{subequations}
The corresponding metasurface exhibits birefringence, since it transforms the two polarizations differently.

\subsubsection{Quadruple Transformation}

\begin{figure}[!h]
    \centering
         \includegraphics[width=1\linewidth, trim={-0.1in 0in -0.4in 0in}]{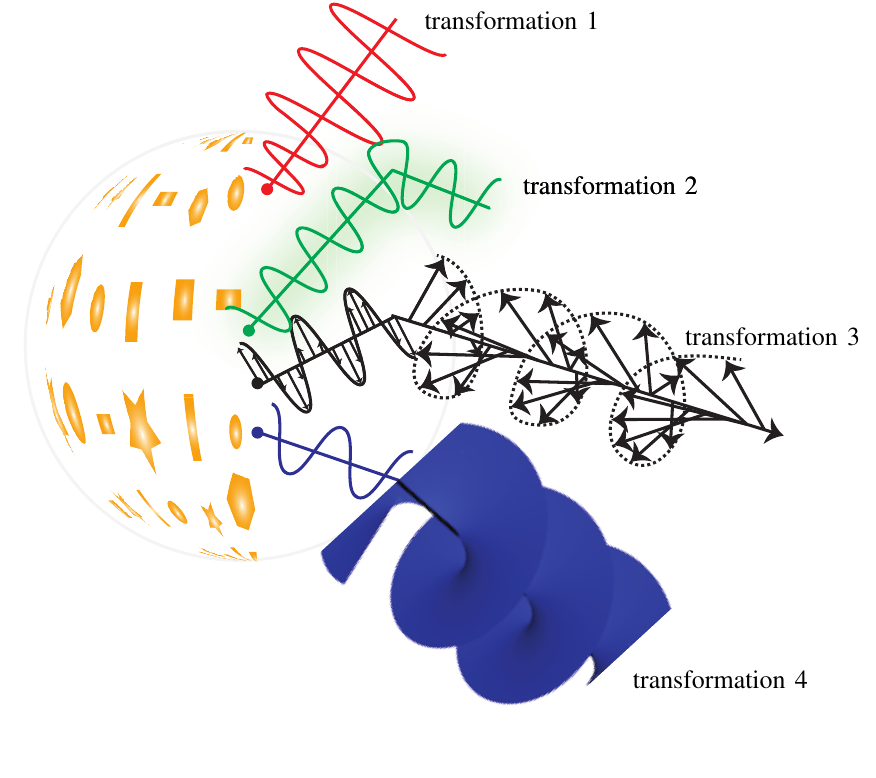}{
        }
        \caption{Illustration of a quadruple transformation, whereby the metasurface simultaneously and independently manipulates waves generated by four different sources. In this particular example, the four transformations are: 1)~amplification, 2)~refraction, 3)~linear to circular polarization transformation, and 4)~zero to nonzero orbital angular momentum transformation.}
   \label{FIG:quadrupletransform}
\end{figure}

In the absence of any constraints, and particularly without requiring~\eqref{reciprocity} to~\eqref{non-gyrotropy} -- i.e. having a loss/gain, nonreciprocal and gyroropic structure -- the spherical metasurface may achieve any arbitrary quadruple transformation\footnote{In this case, a mathematical solution for the bianisotropic susceptibility tensor functions is always be found, but these mathematical functions will not necessarily be \emph{practically} realizable. For instance, if the transformation is too drastic, the corresponding susceptibility functions may exhibit spatial variations that are too high for sampling by typical $p\approx\lambda/5$ particles. Or the required gain or/and nonreciprocity requirements may be unrealizable in the available technology, etc.}, as illustrated in \figref{FIG:quadrupletransform}.

In this case, Eqs~\eqref{Eq:GSTC_EH} represent a system of 16 equations in 16~unknowns. For instance, the first line of~\eqref{Eq:GSTC_expanded_h} splits into the 4 equations
\begin{small}
\begin{equation}\label{Eq:GSTC_four_matrix}
\begin{split}
     \begin{bmatrix}
   -\triangle H_{\phi 1} \\
   -\triangle H_{\phi 2}\\
      -\triangle H_{\phi 3}\\
         -\triangle H_{\phi 4}
  \end{bmatrix}^\text{T}=j\omega\epsilon_0 \begin{bmatrix}
 \chi^{\theta\theta}_\text{ee}&\chi^{\theta\phi}_\text{ee}
  \end{bmatrix}\begin{bmatrix}
   E_{\theta1,\text{av}} &   E_{\theta2,\text{av}}&   E_{\theta3,\text{av}}&   E_{\theta4,\text{av}}\\
   E_{\phi1,\text{av}}&   E_{\phi2,\text{av}}&   E_{\phi3,\text{av}}&   E_{\phi4,\text{av}}
  \end{bmatrix}\\
  +jk_0 \begin{bmatrix}
\chi^{\theta \theta}_\text{em}&\chi^{\theta \phi}_\text{em}
  \end{bmatrix} \begin{bmatrix}
                  H_{\theta1,\text{av}}&H_{\theta2,\text{av}}&H_{\theta3,\text{av}}&H_{\theta4,\text{av}} \\
                  H_{\phi1,\text{av}}&H_{\phi2,\text{av}}&H_{\phi3,\text{av}}&H_{\phi4,\text{av}}
                \end{bmatrix},
\end{split}
\end{equation}
\end{small}
whose solution is
\begin{small}
\begin{equation}\label{four_chi}
 \begin{bmatrix}
   j\omega\epsilon_0 \chi^{\theta\theta}_\text{ee}\\
 j\omega\epsilon_0\chi^ {\theta\phi}_\text{ee}\\
      jk_0\chi^{\theta \theta}_\text{em}\\
         jk_0\chi^{\theta \phi}_\text{em}
  \end{bmatrix}=
       \begin{bmatrix}
  E_{\theta1,\text{av}}& E_{\phi1,\text{av}}&H_{\theta1,\text{av}}&H_{\phi1,\text{av}}\\
  E_{\theta2,\text{av}}& E_{\phi2,\text{av}}&H_{\theta2,\text{av}}&H_{\phi2,\text{av}} \\
  E_{\theta3,\text{av}}& E_{\phi3,\text{av}}&H_{\theta3,\text{av}}&H_{\phi3,\text{av}} \\
     E_{\theta4,\text{av}}& E_{\phi4,\text{av}}&H_{\theta4,\text{av}}&H_{\phi4,\text{av}} \\
  \end{bmatrix}^{-1}
     \begin{bmatrix}
   -\triangle H_\phi1 \\
   -\triangle H_\phi2\\
      -\triangle H_\phi3\\
         -\triangle H_\phi4
  \end{bmatrix}.
\end{equation}
\end{small}

\subsection{Scattering Parameter Mapping}\label{sec:chi_to_RT}

\begin{figure}[!h]
    \centering
         \includegraphics[width=1\linewidth, trim={-0.1in 0in -0.4in 0in}]{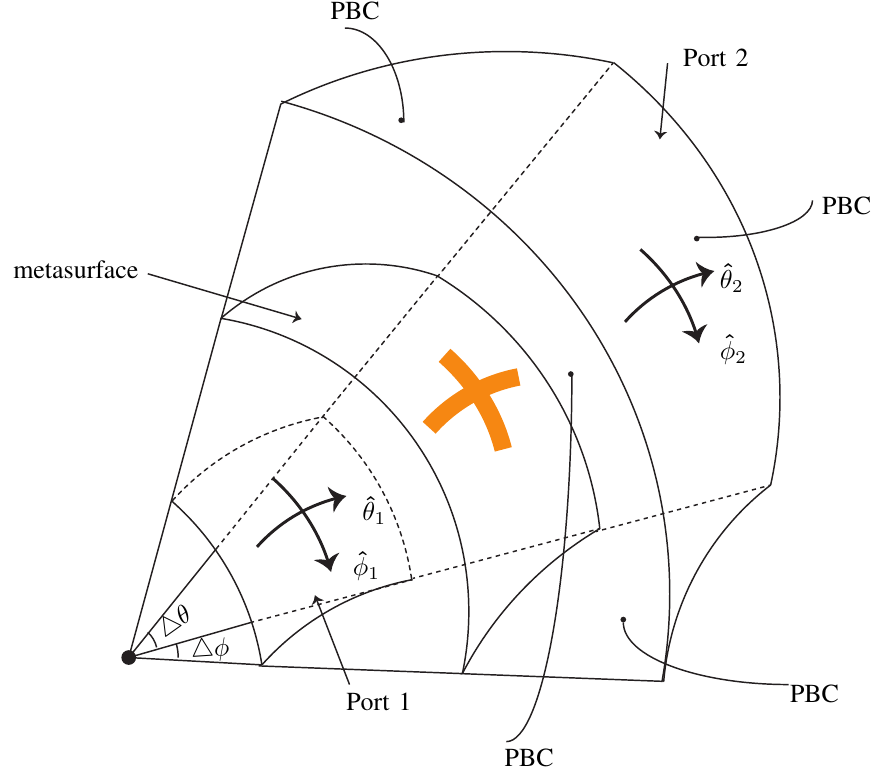}{
        }
        \caption{Unit cell spherical-cap port configuration for the scattering parameter mapping operation in the synthesis procedure.}
   \label{FIG:PBC}
\end{figure}

In the holistic metasurface synthesis procedure described in Sec.~\ref{sec:intro}~\cite{YV_Computation_MS2016,2017_12_Achouri_Nanophotonics}, the susceptibility synthesis operation is followed by the determination of the physical metasurface structure via scattering parameter mapping. In the present case of a spherical metasurface, this mapping is different to that for the planar metasurface due to the different geometry. Instead considering wave scattering between two planar ports, we need to consider here scattering between two spherical-cap ports, as shown in \figref{FIG:PBC}. We will here only present the scattering parameter mapping procedure, without any specific physical metasurface design, which will be presented elsewhere with experimental results.

The scattering parameter mapping method consists in the following steps, followed in~\cite{synthesis_planar_KA2015} for the case of planar metasurface:
\begin{enumerate}
  \item discretize the synthesized spherical susceptibility functions into subwavelength spherical-cap unit cells, as shown in \figref{FIG:PBC}, typically of size (lattice period) in the order of $p\approx\lambda/5$;
  \item for each unit cell, select a scattering particle geometry that is physically consistent with the synthesized susceptibility at the corresponding point\footnote{For instance, if the susceptibility at that point is non-gyrotropic and if one uses conducting particles, the particles should not include asymmetric bends, and one may then choose a straight cross or a Jerusalem cross, while if the susceptibility is chiral, one may choose a gammadion cross.};
  \item compute the scattering parameters of that unit-cell, within periodic boundary conditions (PBCs) to approximate coupling between smoothly varying unit cells (assuming proper sampling), and between the two spherical-cap ports shown in \figref{FIG:PBC} for polarizations corresponding to the tensorial nature of the synthesized susceptibility;
  \item convert the resulting unit-cell scattering parameter functions to the corresponding susceptibility functions, as will be shown next by an example, and compare these functions with the synthesized susceptibility functions.
  \item adjust the geometrical parameters of the physical unit cell until its susceptibility functions match the synthesized ones, and repeat this operation for all the unit cells;
  \item combine the so-designed unit cells to form the final spherical metasurface structure.
\end{enumerate}

For instance, consider the monoanisotropic transformation in Sec.~\ref{sec:monoanis_transf}, whose susceptibility functions are given by~\eqref{GSTC_X_e}. After following steps~1) to~3) above, one needs to establish the proper conversion formulas for step~4). For this purpose, due to the absence of gyrotropy, we only need to consider the uncoupled orthogonal $\theta$ and $\phi$ ports, with reflection and transmission parameters $R_\theta=E_\theta^\text{r}/E_\theta^\text{i}$ ($S_{11}$ for $\theta$ input port), $T_\theta=E_\theta^\text{t}/E_\theta^\text{i}$ ($S_{21}$ for $\theta$ input and output ports), $R_\phi=E_\phi^\text{r}/E_\phi^\text{i}$ ($S_{11}$ for $\phi$ input port) and $T_\phi=E_\phi^\text{t}/E_\phi^\text{i}$ ($S_{21}$ for $\phi$ input and output ports)\footnote{If the two components were coupled, then one would need to perform extra scattering parameter computations, involving for instance $T_{\theta\phi}=E_\theta^\text{t}/E_\phi^\text{i}$, i.e. $S_{21}$ for $\phi$ input port and $\theta$ output port.}. The sought after relations are found upon specifying the difference and average fields in terms of those parameters, for instance $\triangle H_\phi=T_\theta H_\phi^\text{i}-(1-R_\theta)H_\phi^\text{i}$ and  $E_{\theta,\text{av}}=[T_\theta E_\theta^\text{i}+(1+R_\theta)E_\theta^\text{i}]/2$, into~\eqref{GSTC_X_e}, which yields
\begin{subequations}\label{Eq:chisacteering}
\begin{align}
\chi_\text{ee}^{\theta\theta}=\frac{-2[T_\theta-(1-R_\theta)]}{j\omega\epsilon_0\eta(T_\theta+1+R_\theta)},\\
\chi_\text{ee}^{\phi\phi}=\frac{2[T_\phi-(1-R_\phi])}{j\omega\epsilon_0\eta(T_\phi+1+R_\phi)},\\
\chi_\text{mm}^{\theta\theta}=\frac{2\eta[T_\phi-(1+R_\phi)]}{j\omega\mu_0(T_\phi+1-R_\phi)},\\
\chi_\text{mm}^{\phi\phi}=\frac{-2\eta[T_\theta-(1+R_\theta)]}{j\omega\mu_0(T_\theta+1-R_\theta)}.
\end{align}
\end{subequations}
From this point, one proceeds to steps 5)~ and~6) above, which completes the synthesis.

\section{Illustrative Examples}\label{sec:Ill_ex}

In this section, we illustrate the spherical metasurface synthesis presented in Sec.~\ref{sec:synthesis} with the help of three examples in two steps:
\begin{itemize}
  \item First, we \emph{compute the susceptibility functions}~\eqref{Eq:tensors} corresponding to the specified fields using the general equations~\eqref{Eq:GSTC_EH}, which is the essence of the \emph{synthesis} procedure.
  \item Second, we \emph{validate this synthesis} by comparing the fields scattered by the metasurfaces with such susceptibilities with the specified fields.
\end{itemize}

The latter, which is an \emph{analysis} operation, is performed by modelling the spherical surface susceptibility functions by volume-diluted susceptibility functions, following the procedure described in Appendix~\ref{sec:sur_vol_chi}, in the full-wave commercial full-wave FEM-based software COMSOL.

The three examples will share the following features:

\begin{itemize}
  \item Since COMSOL does not support bianisotropic media, i.e. assumes $\te{\chi}_\text{em}=\te{\chi}_\text{me}=0$, the metasurface will be monoanisotropic\footnote{We could naturally still have plotted the susceptibilities for such media, but these functions would not be very informative.}.
  \item Since COMSOL requires excessive memory for 3D simulations, the metasurface will have variations only along one direction, corresponding to a transformation that is symmetric in the other direction. Specifically, the metasurface will belong to the category (c) in \figref{FIG:class} (equivalent to category (b) upon $\pi/2$ rotation).
  \item The metasurface will be reflection-less, according to assumption b) in the second paragraph of Sec.~\ref{sec:problem}.
  \item For simplicity, and without loss of generality, the metasurface will be surrounded by vacuum ($\epsilon_1=\epsilon_2=\epsilon_0$ and $\mu_1=\mu_2=\mu_0$).
  \item The sources will be infinitesimal vertical dipoles.
  \item Finally, the transformations will be specified in the far field.
\end{itemize}

\subsection{Illusion Transformation}

The first example is about illusion transformation. Specifically, the spherical metasurface is required to transform the field radiated by an off-centered source into the field produced by a virtual centered source.

Assuming the source location $(x,y,z)=(0,0,z_0)$, the corresponding incident field specification is
\begin{subequations}\label{Eq:case1_in}
\begin{align}
  E_{\theta}^\text{i}|_{r=a^-} & =-\frac{I\ell}{j\omega \epsilon_0}\frac{e^{-jk_0\sqrt{(a^-\sin\theta)^2+(a^-\cos\theta-z_0)^2}}}{4\pi\sqrt{(a^-\sin\theta)^2+(a^-\cos\theta-z_0)^2}}k_0^2\sin\theta,\\
  H_{\phi}^\text{i}|_{r=a^-} & =I\ell\frac{e^{-jk_0\sqrt{(a^-\sin\theta)^2+(a^-\cos\theta-z_0)^2}}}{4\pi\sqrt{(a^-\sin\theta)^2+(a^-\cos\theta-z_0)^2}}jk_0\sin\theta ,
\end{align}
\end{subequations}
while the transmitted field specification is
\begin{subequations}\label{Eq:case1_out}
\begin{align}
  E_{\theta}^\text{t}|_{r=a^+} & =-T\frac{I\ell}{j\omega \epsilon_0}\frac{e^{-jk_0a^+}}{4\pi a^+}k_0^2\sin\theta,\\
  H_{\phi}^\text{t}|_{r=a^+} & =TI\ell\frac{e^{-jk_0a^+}}{4\pi a^+}jk_0\sin\theta,
\end{align}
\end{subequations}
where $I\ell$ is the dipole moment ($I$: current, $\ell$: length) and $k_0=\omega/c$ is the free-space wavenumber ($\omega$: angular frequency, $c$: velocity of light in vacuum) and $T$ is the transmission coefficient. The incident and transmitted fields are related by the condition of local power conservation~\cite{GL_perfect_refraction2017},
\begin{equation}\label{Eq:case1_energy_cons}
\frac{1}{2}\text{Re}[(E_{\theta}^\text{i}\times  H_{\phi}^{\text{i},*})]=\frac{1}{2}\text{Re}[(E_{\theta}^\text{t}\times  H_{\phi}^{\text{t},*})],
\end{equation}
which sets the transmission coefficient to
\begin{equation}\label{Eq:case1_energy_T}
T=\frac{a^+}{\sqrt{(a^-\sin\theta)^2+(a^-\cos\theta-z_0)^2}}.
\end{equation}

\begin{figure}[!h]
    \centering
         \includegraphics[width=1\linewidth, trim={-0.1in 0in -0.4in 0in}]{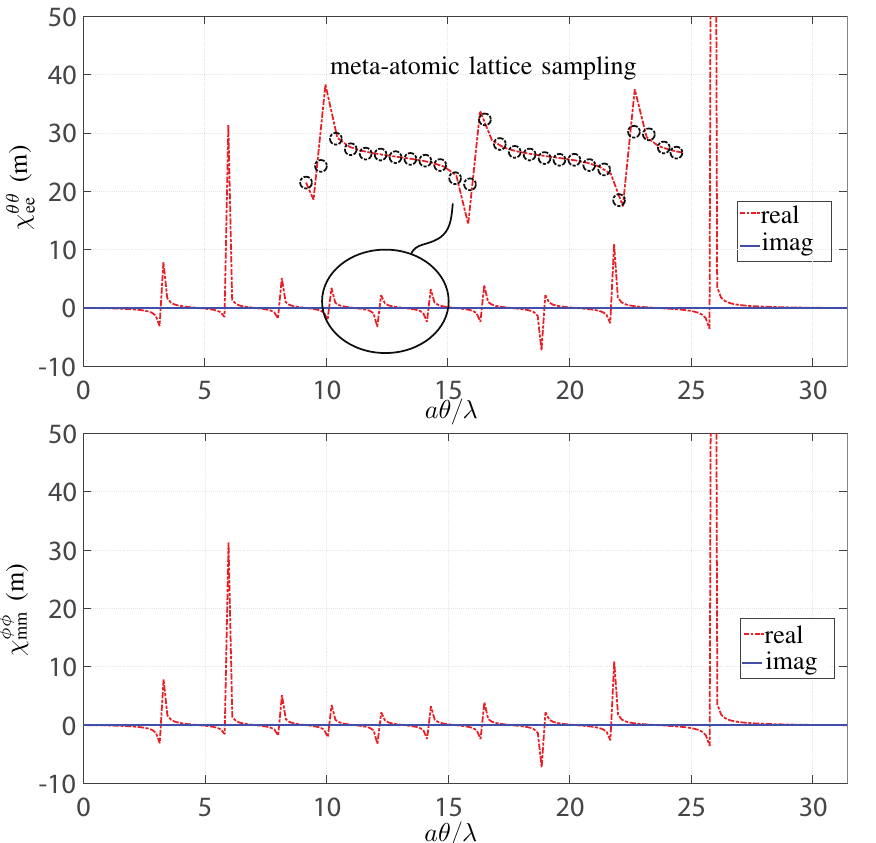}{
        }
        \caption{Susceptibility functions for the illusion transformation metasurface example with the parameters $a=10\lambda$ and $z_0=5\lambda$. The normalized horizontal axis spans the entire elevation space, i.e. extends from $\theta=0$ to $\theta=\pi$. The inset of the top figure shows the sampling of the susceptibility function with typical $\lambda/5$ scattering particles or periodic unit cells.}
   \label{FIG:offtocenterchi1}
\end{figure}

Inserting these field specifications into~\eqref{GSTC_X_e} yields the susceptibilities $\chi_\text{ee}^{\theta\theta}=\chi_\text{mm}^{\phi\phi}$\footnote{The equality is a result of the reflection-less specification, corresponding to electric and magnetic polarization currents canceling out at the input side (Huygens metasurface).} plotted\footnote{The resulting relations are naturally closed-form expressions, but we do not give them here for the sake of brevity} in \figref{FIG:offtocenterchi1} and $\chi_\text{ee}^{\phi\phi}=\chi_\text{mm}^{\theta\theta}=0$. The imaginary parts of susceptibilities are zero, which indicates that the metasurface performing the specified transformation is loss-less and gain-less. We see, with the help of the lattice in the inset of the top figure, that in this design, typical $\lambda/5$ particles or cells~\cite{2017_12_Achouri_Nanophotonics} can hardly sample the required susceptibility, except on the smooth parts of it. This issue may be resolved by increasing the electrical size the sphere or reduce the distance of the source to the center, if this is acceptable.

\begin{figure}[!h]
    \centering
         \includegraphics[width=1\linewidth, trim={-0.1in 0in -0.4in 0in}]{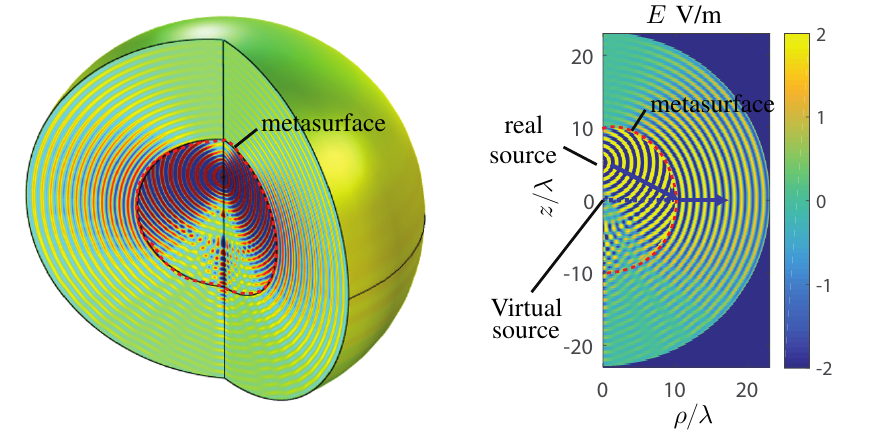}{
        }
        \caption{Full-wave validation of the illusion transformation accomplished by the metasurface with susceptibilities plotted in \figref{FIG:offtocenterchi1}. The 3D picture is obtained from revolving the 2D-computed fields about the $z$ axis.}
   \label{FIG:offtocenter}
\end{figure}

Finally, \figref{FIG:offtocenter} provides the full-wave validation of the metasurface synthesis, where the wave scattered from the metasurface clearly seems to be radiated by centered (virtual) source.

\subsection{Ring Focusing}

\begin{figure}[!h]
    \centering
         \includegraphics[width=1\linewidth, trim={-0.1in 0in -0.4in 0in}]{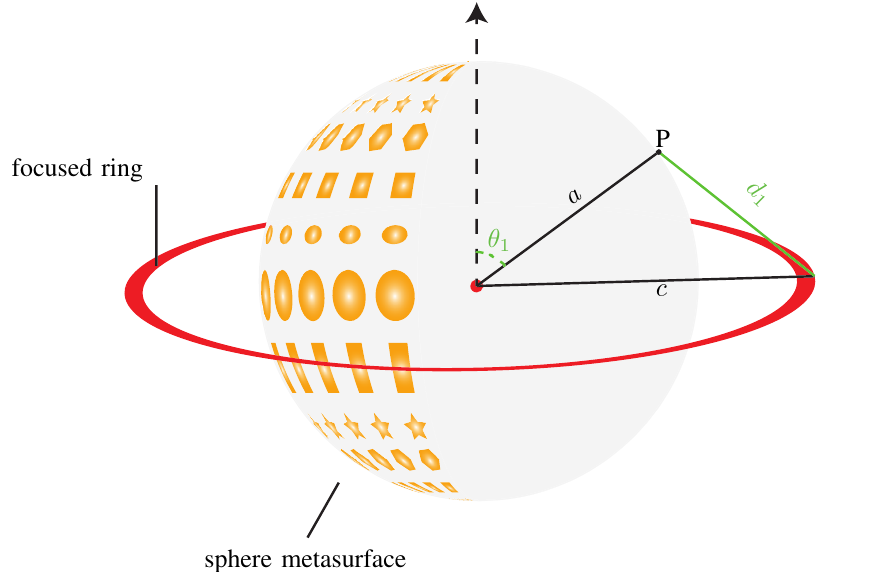}{
        }
        \caption{Focussing of the wave radiated by a centered point source on a ring of radius $c$.}
   \label{FIG:satune}
\end{figure}

The second example is a metasurface focusing the field radiated by a centered source onto a ring, as illustrated in \figref{FIG:satune}. For such focusing, the total optical path from the source to the ring via any point $P$ on the metasurface should be constant, namely
\begin{subequations}\label{Eq:phase}
\begin{equation}
\begin{split}
-jk_0a+\Phi_P-jk_0d(\theta)
&=-jk_0[a+d(\theta)]+\Phi_P\\
&=\text{const.},
\end{split}
\end{equation}
with
\begin{equation}
d(\theta)=\sqrt{a^2+c^2-2ac\sin(\theta)},
\end{equation}
\end{subequations}
where $\Phi_P(\theta)=jk_0[a+d(\theta)]+\text{const.}$ corresponds to the correction phase function to be provided by the metasurface.

The corresponding incident and transmitted field specifications are
\begin{subequations}\label{Eq:saturn_field_in}
\begin{align}
  E_{\theta}^\text{i}|_{r=a^-} & =-\frac{I\ell}{j\omega \epsilon_0}\frac{e^{-jk_0a^-}}{4\pi a^-}k_0^2\sin(\theta),\\
  H_{\phi}^\text{i}|_{r=a^-} & =I\ell\frac{e^{-jk_0a^-}}{4\pi a^-}jk_0\sin(\theta),
\end{align}
\end{subequations}
and
 \begin{subequations}\label{Eq:saturn_field_out}
\begin{align}
  E_{\theta}^\text{t}|_{r=a^+} & =T\eta e^{jk_0d(\theta)},\\
  H_{\phi}^\text{t}|_{r=a^+} & =Te^{jk_0d(\theta)},
\end{align}
\end{subequations}
where $T$ is found from~\eqref{Eq:case1_energy_cons} as
\begin{equation}
T=\frac{I\ell}{4\pi a^-}k_0\sin\theta.
\end{equation}

\begin{figure}[!h]
    \centering
         \includegraphics[width=1\linewidth, trim={-0.1in 0in -0.4in 0in}]{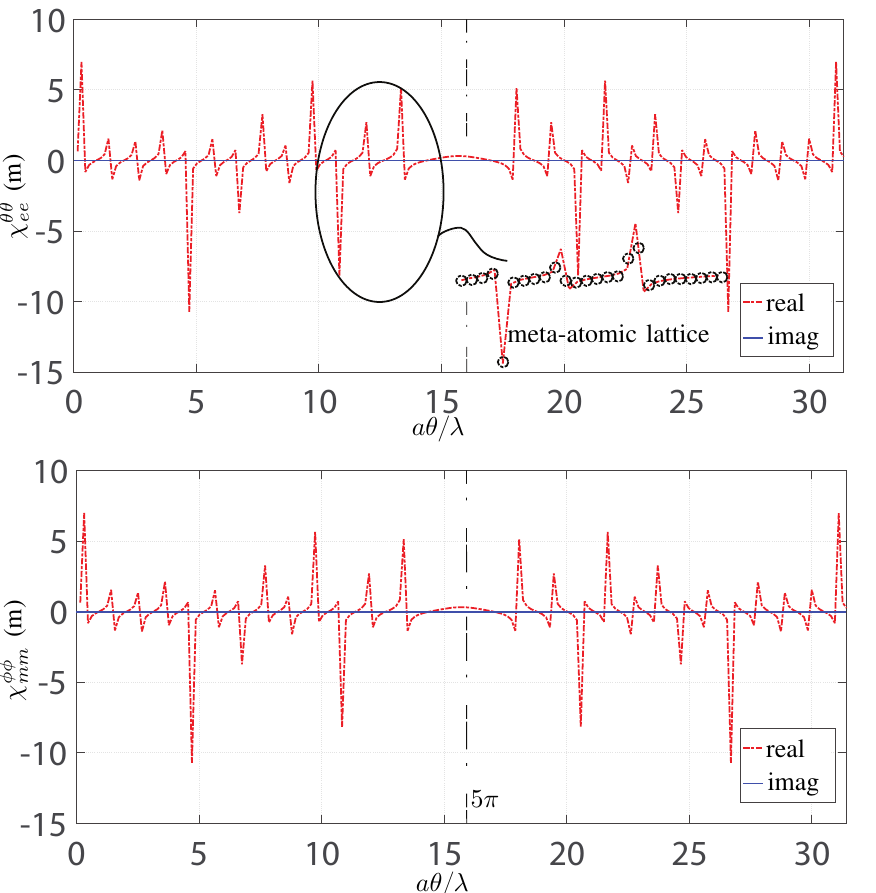}{
        }
        \caption{Susceptibility functions for the ring focusing transformation illustrated in \figref{FIG:satune} with the same parameters as in \figref{FIG:offtocenterchi1}.}
   \label{FIG:pointtoringchi1}
\end{figure}

Inserting these field specifications into~\eqref{GSTC_X_e} yields the susceptibilities $\chi_\text{ee}^{\theta\theta}=\chi_\text{mm}^{\phi\phi}$ plotted in \figref{FIG:pointtoringchi1} and $\chi_\text{ee}^{\phi\phi}=\chi_\text{mm}^{\theta\theta}=0$.
As expected from the symmetry of the transformation, these susceptibility functions are symmetry about the equator of the metasurface, i.e. at $\theta=\pi/2$ or $a\theta/\lambda=5\pi$. Moreover, the susceptibilities are minimal in the vicinity of the equator where the required transformation is minimal given the doughnut radiation pattern of the vertical dipole source. Similar considerations is in the previous examples may be made about sampling.

\begin{figure}[!h]
    \centering
         \includegraphics[width=1\linewidth, trim={-0.1in 0in -0.4in 0in}]{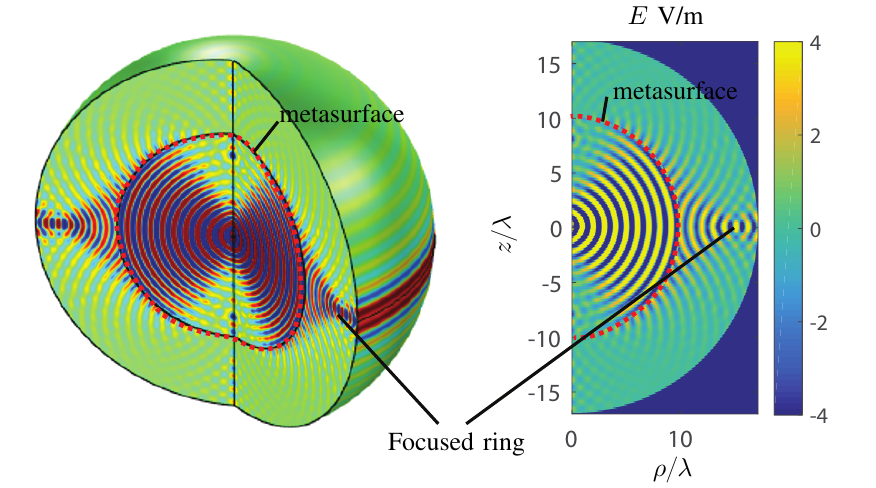}{
        }
        \caption{Full-wave validation of the ring focusing accomplished by the metasurface with the susceptibilities plotted in \figref{FIG:pointtoringchi1}.}
   \label{FIG:pointtoring}
\end{figure}

Finally, \figref{FIG:pointtoring} provides the full-wave validation of the metasurface synthesis, where the wave scattered from the metasurface clearly focusses on the specified ring region.

\subsection{Birefringence}

\begin{figure}[!h]
    \centering
         \includegraphics[width=1\linewidth, trim={-0.1in 0in -0.4in 0in}]{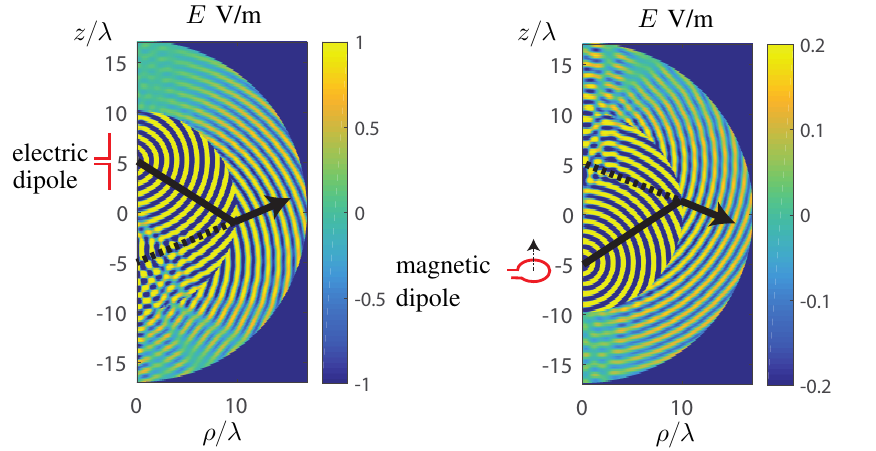}{
        }
        \caption{Full-wave description and validation, for the susceptibilities plotted in \figref{FIG:doublechi}, of a birefringent (double) transformation with electric and magnetic sources placed at $(x,y,z)=(0,0,z_1=5\lambda$) and $(x,y,z)=(0,0,z_2=-5\lambda$), respectively, and $a=10\lambda$.}
   \label{FIG:double}
\end{figure}

The third example is a birefringent (double-transformation) metasurface transforming the fields radiated by two off-centered orthogonal electric and magnetic sources into the fields of virtual sources of the same nature place at the position of the other source, as illustrated in \figref{FIG:double}.

The fields corresponding to this transformations are
\begin{subequations}\label{Eq:double_field_in1}
\begin{align}
  E_{\theta1} ^\text{i}|_{r=a^-}& =I\ell jk_0\eta\frac{e^{-jk_0\sqrt{(a^-\sin\theta)^2+(a^-\cos\theta-z_1)^2}}}{4\pi \sqrt{(a^-\sin\theta)^2+(a^-\cos\theta-z_1)^2}}\sin(\theta),\\
  H_{\phi1}^\text{i}|_{r=a^-} & =I\ell jk_0\frac{e^{-jk_0\sqrt{(a^-\sin\theta)^2+(a^-\cos\theta-z_1)^2}}}{4\pi \sqrt{(a^-\sin\theta)^2+(a^-\cos\theta-z_1)^2}}\sin(\theta),
\end{align}
\end{subequations}
\begin{subequations}\label{Eq:double_field_in2}
\begin{align}
  E_{\phi2}^\text{i}|_{r=a^-} & =IA\eta k_0^2\frac{e^{-jk_0\sqrt{(a^-\sin\theta)^2+(a^-\cos\theta-z_2)^2}}}{4\pi \sqrt{(a^-\sin\theta)^2+(a^-\cos\theta-z_2)^2}}\sin(\theta),\\
  H_{\theta2}^\text{i} |_{r=a^-}& =-IAk_0^2\frac{e^{-jk_0\sqrt{(a^-\sin\theta)^2+(a^-\cos\theta-z_2)^2}}}{4\pi\sqrt{(a^-\sin\theta)^2+(a^-\cos\theta-z_2)^2}}\sin(\theta)
\end{align}
\end{subequations}
and
\begin{subequations}\label{Eq:double_field_out1}
\begin{align}
  E_{\theta1}^\text{t}|_{r=a^+}& =T_1I\ell\eta\frac{e^{-jk_0\sqrt{(a^+\sin\theta)^2+(a^+\cos\theta-z_2)^2}}}{4\pi \sqrt{(a^+\sin\theta)^2+(a^+\cos\theta-z_2)^2}}jk_0\sin(\theta),\\
  H_{\phi1}^\text{t}|_{r=a^+}& =T_1I\ell\frac{e^{-jk_0\sqrt{(a^+\sin\theta)^2+(a^+\cos\theta-z_2)^2}}}{4\pi \sqrt{(a^+\sin\theta)^2+(a^+\cos\theta-z_2)^2}}jk_0\sin(\theta),
\end{align}
\end{subequations}
\begin{subequations}\label{Eq:double_field_out2}
\begin{align}
  E_{\phi2}^\text{t}|_{r=a^+}& =T_2IA\eta k_0^2\frac{e^{-jk_0\sqrt{(a^-\sin\theta)^2+(a^-\cos\theta-z_1)^2}}}{4\pi \sqrt{(a^-\sin\theta)^2+(a^-\cos\theta-z_1)^2}}\sin(\theta),\\
  H_{\theta2}^\text{t}|_{r=a^+}& =-T_2IAk_0^2\frac{e^{-jk_0\sqrt{(a^-\sin\theta)^2+(a^-\cos\theta-z_1)^2}}}{4\pi\sqrt{(a^-\sin\theta)^2+(a^-\cos\theta-z_1)^2}}\sin(\theta)
\end{align}
\end{subequations}
with
\begin{subequations}
\begin{align}
T1=\sqrt{\frac{(a^+\sin\theta)^2+(a^+\cos\theta-z_2)^2}{a^-\sin\theta)^2+(a^-\cos\theta-z_1)^2}},\\
T2=\sqrt{\frac{(a^+\sin\theta)^2+(a^+\cos\theta-z_1)^2}{a^-\sin\theta)^2+(a^-\cos\theta-z_2)^2}}.
\end{align}
\end{subequations}

\begin{figure}[!h]
    \centering
         \includegraphics[width=1\linewidth, trim={-0.1in 0in -0.4in 0in}]{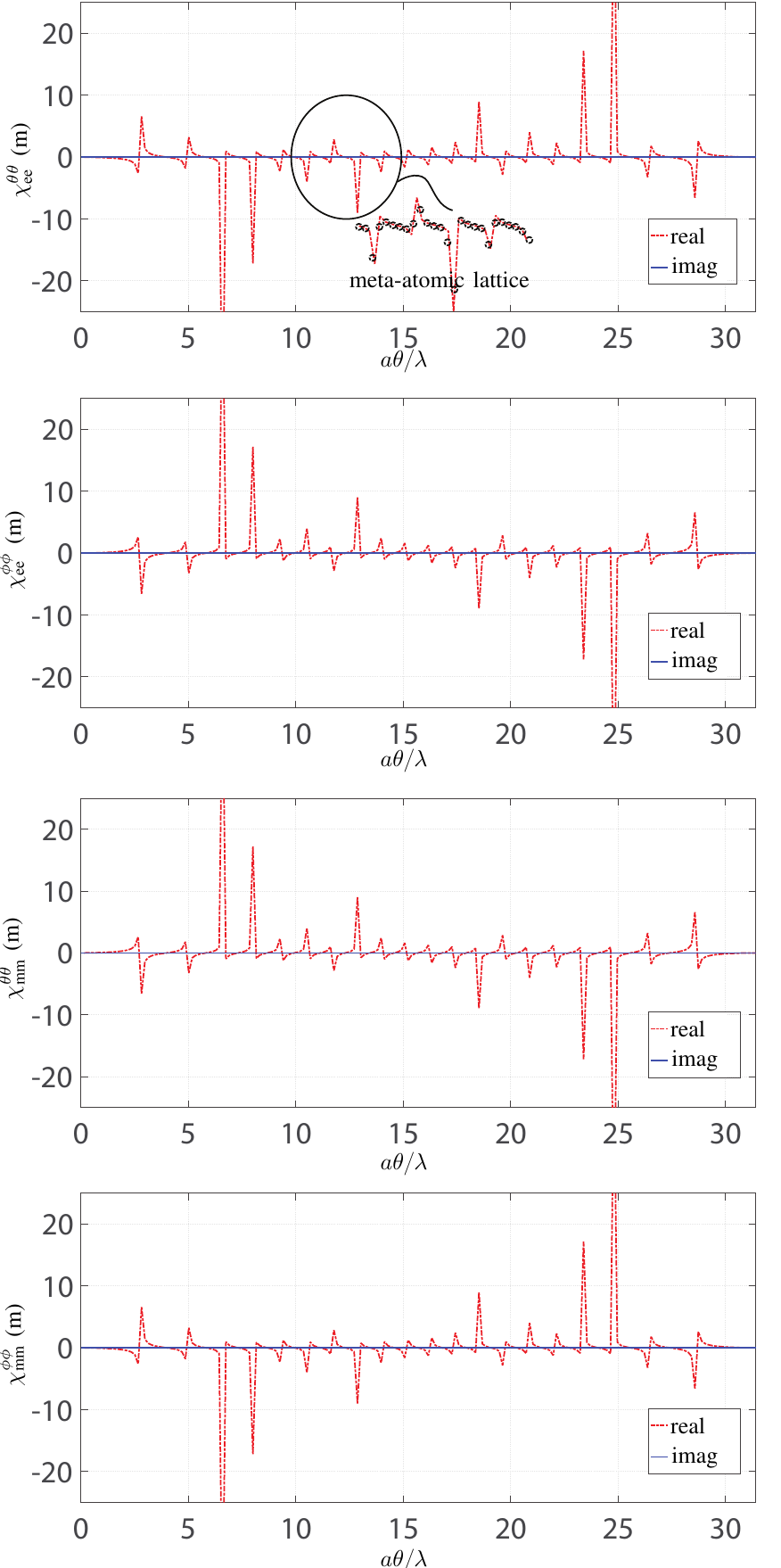}{
        }
        \caption{Susceptibility functions for the metasurface in \figref{FIG:double}.}
   \label{FIG:doublechi}
\end{figure}

Inserting these field specifications into~\eqref{Eq:doublechi} yields the four susceptibility functions plotted in \figref{FIG:doublechi}, whose symmetry is again expected.

The full-wave simulation results, in \figref{FIG:double}, are in perfect agreement with the expectation.

\section{Conclusion}
\label{sec:concl}

This paper has extended the susceptibility-GSTC synthesis of planar metasurfaces to spherical metasurfaces. In contrast to the cylindrical metasurface that has been the first non-planar metasurface modeled by susceptibility-GSTC, the spherical metasurface is has a non-zero intrinsic curvature and hence exhibits particularly interesting characteristics, some of which have been pointed out. The paper paves the way for the study of other canonical-shape metasurfaces of non-zero intrinsic curvature, and prompts for the exploration of irregular shaped metasurfaces combining GSTCs with conformal mapping techniques.

\bibliographystyle{IEEEtran}
\bibliography{SPHERICAL_SYN}

\appendix
\subsection{GSTCs Derivation}\label{GSTC_Deriv}

Since GSTCs are \emph{local}, they are the same for curved metasurfaces, including the spherical metasurface of interest here, as for planar metasurface\footnote{This is naturally under the assumption of Footnote~\ref{footnote:Rayleigh}.}. They may be derived in different fashions. The most rigorous one, allowing for any discontinuity order, is given by Idemen in~\cite{GSTCs_Idemen2011} and clarified in the appendix of~\cite{synthesis_planar_KA2015}. However, this approach is mostly mathematical and does not clearly reveal how to take into account the (possibly different) media surrounding the metasurface. An alternative approach, more physical and classical, is given in~\cite{tretyakov2016framework} for plane waves. We present here a derivation that is in the vein of~\cite{tretyakov2016framework}, but that is more general, starting from Maxwell and constitutive relations, involving both volume and surface polarization densities, and applying to any type of waves.

Assuming time-harmonic ($e^{+j\omega t}$) waves, \emph{symmetric} Maxwell equations take the form
\begin{subequations}\label{Eq:harm_maxwell}
\begin{align}
\nabla\times\ve{E}&=-j\omega\ve{B}-\ve{K}_\text{imp}, \\
\nabla\times\ve{H}&=j\omega\ve{D}+\ve{J}_\text{imp},
\end{align}
\end{subequations}
where $\ve{J}_\text{imp}$ (A/m$^2$) is the impressed electric current density and $\ve{K}_\text{imp}$ (V/m$^2$) is the (fictitious) impressed magnetic current density, and where the fields and  are related to the electric polarization density $\ve{P}$ and the magnetic polarization density $\ve{M}$ by the constitutive relations
\begin{subequations}\label{eq:const_rel}
\begin{equation}\label{eq:const_rel_el}
\ve{D}=\epsilon_0\ve{E}+\ve{P},
\quad\text{or}\quad
\ve{E}=(\ve{D}-\ve{P})/\epsilon_0,
\end{equation}
\begin{equation}\label{eq:const_rel_mag}
\ve{B}=\mu_0(\ve{H}+\ve{M}),
\quad\text{or}\quad
\ve{H}=\ve{B}/\mu_0-\ve{M},
\end{equation}
\end{subequations}
with $\ve{P}$ and $\ve{M}$ measured in C/m$^2$ and A/m, respectively.

The essence of GSTCs is to model the metasurface as a thin sheet of equivalent polarization currents. Therefore, the fields $\ve{D}$ and $\ve{E}$ have to be written in terms of $\ve{P}$, as in~\eqref{eq:const_rel_el}, and $\ve{B}$ and $\ve{H}$ have to be written in terms of $\ve{M}$, as in~\eqref{eq:const_rel_mag}. Inserting these relations into~\eqref{Eq:harm_maxwell} yields
\begin{subequations}\label{Eq:max_cons}
\begin{align}
\nabla\times[(\ve{D}-\ve{P})/\epsilon_0)]&=-j\omega\mu_0(\ve{H}+\ve{M})-\ve{K}_\text{imp},\\
\nabla\times(\ve{B}/\mu_0-\ve{M})&=j\omega(\epsilon_0\ve{E}+\ve{P})+\ve{J}_\text{imp}.
\end{align}
\end{subequations}

Assuming first-order metasurface discontinuity~\cite{synthesis_planar_KA2015}, the polarization densities decompose into volume and surface parts as
\begin{subequations}\label{Eq:polar_sur_vol}
\begin{align}
\ve{P}&=\ve{P}_\text{v}+\ve{P}_\text{s}\delta({r}),\\
\ve{M}&=\ve{M}_\text{v}+\ve{M}_\text{s}\delta({r}).
\end{align}
\end{subequations}
Inserting~\eqref{Eq:polar_sur_vol} into~\eqref{Eq:max_cons}, and transferring the surface parts to the right-hand sides, yields
\begin{subequations}\label{Eq:app_max_cons_EH}
\begin{align}
\begin{split}
\nabla\times\left(\frac{\ve{D}-\ve{P}_\text{v}}{\epsilon_0}\right)&=-j\omega\mu_0\ve{H}-j\omega\mu_0\ve{M}_\text{v}\\
-&j\omega\mu_0\ve{M}_\text{s}+\nabla\times\left[\frac{\ve{P}_\text{s}\delta(r)}{\epsilon_0}\right]-\ve{K}_\text{imp},
\end{split}\\
\begin{split}
\nabla\times\left(\frac{\ve{B}}{\mu_0}-\ve{M}_\text{v}\right)&=j\omega\epsilon_0\ve{E}+j\omega\epsilon_0\ve{P}_\text{v}\\
+&j\omega\epsilon_0\ve{P}_\text{s}+\nabla\times[\ve{M}_\text{s}\delta(r)]+\ve{J}_\text{imp}.
\end{split}
\end{align}
\end{subequations}

The metasurface discontinuity may now be analyzed by integrating~\eqref{Eq:app_max_cons_EH} over the usual rectangular surface around the interface between the two media, that supports here the metasurface, as shown in~\figref{FIG:curvedboundary}, and applying Stokes theorem. This yields
\begin{subequations}\label{Eq:app_max_cons_int}
\begin{align}
\begin{split}
\oint \left(\frac{\ve{D}-\ve{P}_\text{v}}{\epsilon_0}\right)\cdot d\ve{l}=-j\omega\mu_0\iint(\ve{H}+\ve{M}_\text{v})\cdot d\ve{S}\\
-j\omega\mu_0\iint\delta(r)\ve{M}_\text{s}\cdot d\ve{S} +\oint\left[\frac{\delta(r)\ve{P}_\text{s}}{\epsilon_0}\right]\cdot d\ve{l}-\iint\ve{K}_\text{imp}\cdot d\ve{S},
\end{split}\\
\begin{split}
\oint  \left(\frac{\ve{B}}{\mu_0}-\ve{M}_\text{v}\right)\cdot d\ve{l}=j\omega\epsilon_0\iint(\ve{E}+\ve{P}_\text{v})\cdot d\ve{S}\\
+j\omega\epsilon_0\iint\delta(r)\ve{P}_\text{s}\cdot d\ve{S}+\oint[\ve{M}_\text{s}\delta(r)]\cdot d\ve{l}+\iint\ve{J}_\text{imp}\cdot d\ve{S}.
\end{split}
\end{align}
\end{subequations}

\begin{figure}[!h]
    \centering
         \includegraphics[width=1\linewidth, trim={-0.1in 0in -0.4in 0in}]{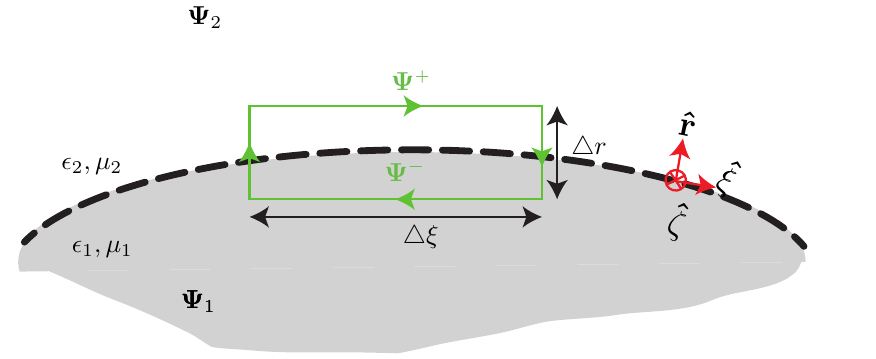}{
         }
        \caption{General curved boundary, supporting a metasurface, surrounded by two media with permittivity-permeability pairs $(\epsilon_1,\mu_1)$ and $(\epsilon_2,\mu_2)$, respectively, with local coordinate system $(\xi,\zeta,r)$ and rectangular integration surface for~\eqref{Eq:app_max_cons_EH} with $\ves{\Psi}=\ve{E},\ve{H},\ve{D},\ve{B},\ve{P},\ve{M}$, labeled $-$ at $r=a^-$ (just below the metasurface in medium~1) and $+$ at $r=a^+$ (just above the metasurface in medium~2).}
   \label{FIG:curveboundary}
\end{figure}

The integrands in the left-hand sides of~\eqref{Eq:app_max_cons_int} are, according to~\eqref{eq:const_rel}, nothing but the electric and magnetic fields in the two media, that be simply written as
\begin{equation}
\frac{\ve{D}^\pm-\ve{P}^\pm_\text{v}}{\epsilon_0}=\ve{E}^\pm,\quad \frac{\ve{B}^\pm}{\mu_0}-\ve{M}^\pm_\text{v}=\ve{H}^\pm.
\end{equation}

With this, the $\xi-r$ projection of Eqs.~\eqref{Eq:app_max_cons_int} integrate to
\begin{subequations}\label{Eq:app_max_cons_disc}
\begin{equation}
\begin{split}
(E_\xi^+&-E_\xi^-)\triangle \xi+ (-E_{r,\text{right}}+E_{r,\text{left}})\triangle r\\
=&-j\omega\mu_0(H_\zeta -M_{\text{v},\zeta}) \triangle \xi \triangle r\\
&-j\omega\mu_0M_{\text{s},\zeta}\delta(r)\triangle \xi \triangle r-(P_{\text{s},r,\text{right}}-P_{\text{s},r,\text{left}})\delta(r)\triangle r/\epsilon_0\\
&-K_{\zeta,\text{v,imp}}\triangle \xi \triangle r-K_{\zeta,\text{s,imp}}\delta(r)\triangle \xi \triangle r,
\end{split}
\end{equation}
\begin{equation}
\begin{split}
(H_\xi^+&-H_\xi^-)\triangle \xi+ (-H_{r,\text{right}}+H_{r,\text{left}})\triangle r\\
=&j\omega\epsilon_0(E_\zeta+P_{v,\zeta}) \triangle \xi \triangle r\\
&+j\omega\epsilon_0P_{\text{s},\zeta}\delta(r) \triangle \xi \triangle r-(M_{\text{s},r,\text{right}}-M_{\text{s},r,\text{left}})\delta(r)\triangle r\\
&-J_{\zeta,\text{v,imp}}\triangle \xi \triangle r-J_{\zeta,\text{s,imp}}\delta(r)\triangle \xi \triangle r.
\end{split}
\end{equation}
\end{subequations}
Taking the limit $\triangle r \rightarrow 0$, replacing $\delta(r)\triangle r\rightarrow 1$, and dividing by $\triangle\xi$, Eqs.~\eqref{Eq:app_max_cons_disc} reduce to
\begin{subequations}\label{Eq:app_max_discon}
\begin{equation}
\begin{split}
(E_\xi^+-E_\xi^-)=&-j\omega\mu_0M_{\text{s},\zeta}\\
&-(P_{\text{s},r,\text{right}}-P_{\text{s},r,\text{left}})/ (\epsilon_0\triangle \xi)-K_{\zeta,\text{s,imp}},\\
(H_\xi^+-H_\xi^-)=&j\omega\epsilon_0P_{s,\zeta}\\
&-(M_{\text{s},r,\text{right}}-M_{\text{s},r,\text{right}})/\triangle \xi+J_{\zeta,\text{s,imp}},
\end{split}
\end{equation}
\end{subequations}
which, in the limit $\triangle\xi\rightarrow 0$, may be written as
\begin{subequations}\label{Eq:app_max_bon}
\begin{align}
\triangle E_\xi&=-j\omega\mu_0 M_{\text{s},\zeta}-\frac{\partial (P_{\text{s},r}/\epsilon_0)}{\partial \xi}-K_{\zeta,\text{s,imp}},\\
&\text{with}\quad\triangle E_\xi=E_\xi^+-E_\xi^-,\\
\triangle H_\xi&=j\omega\epsilon_0P_{\text{s},\zeta}-\frac{\partial M_{\text{s},r}}{\partial \xi}+J_{\zeta,\text{s,imp}},\\
&\text{with}\quad\triangle H_\xi=H_\xi^+-H_\xi^-.
\end{align}
\end{subequations}
Similarly, we find for the $\zeta-r$ projection of Eqs.~\eqref{Eq:app_max_cons_int}
\begin{subequations}\label{Eq:app_max_bon2}
\begin{align}
\triangle E_\zeta&=j\omega\mu_0 M_{\text{s},\xi}-\frac{\partial (P_{\text{s},r}/\epsilon_0)}{\partial \zeta}+K_{\xi,\text{s,imp}},\\
&\text{with}\quad\triangle E_\zeta=E_\zeta^+-E_\zeta^-,\\\triangle H_\zeta&=-j\omega\epsilon_0P_{\text{s},\xi}-\frac{\partial M_{\text{s},r}}{\partial \zeta}-J_{\xi,\text{s,imp}},\\
&\text{with}\quad\triangle H_\zeta=H_\zeta^+-H_\zeta^-.
\end{align}
\end{subequations}
Combining~\eqref{Eq:app_max_bon} and~\eqref{Eq:app_max_bon2} finally yields
\begin{subequations}\label{Eq:app_GSTC_two_media}
\begin{align}
\uve{r}\times\triangle\ve{E}&=-j\omega\mu_0\ve{M_{\text{s},\|}}+\nabla_{\|}(P_{\text{s},r}/\epsilon_0)\times\uve{r}-K_{\|,\text{s,imp}},\\
\uve{r}\times\triangle\ve{H}&=j\omega\ve{P_{\text{s},\|}}-\uve{r}\times\nabla_{\|}M_{\text{s},r}+J_{\|,\text{s,imp}},
\end{align}
\end{subequations}
where the symbol $\|$ denotes the metasurface tangential components $\xi$ and $\zeta$ in this appendix. Relations~\eqref{Eq:app_GSTC_two_media} are the final GSTC equations.

To clearly see how Eqs.~\eqref{Eq:app_GSTC_two_media} account for different surrounding media, as in \figref{FIG:curve_boundary}, we rewrite them explicitly as
\begin{subequations}\label{Eq:app_two_media}
\begin{align}
&\uve{r}\times(\ve{E}^+-\ve{E}^-)=-j\omega\mu_0\ve{M_{\text{s},\|}}+\nabla_{\|}(P_{\text{s},r}/\epsilon_0)\times\uve{r}-K_{\|,\text{s,imp}},\\
&\uve{r}\times(\ve{E}^+/\eta_{2,\text{eff}}-\ve{E}^-/\eta_{1,\text{eff}})=j\omega\ve{P_{\text{s},\|}}-\uve{r}\times\nabla_{\|}M_{\text{s},r}+J_{\|,\text{s,imp}},
\end{align}
\end{subequations}
with the \emph{effective impedances} $\eta_{i,\text{eff}}=\eta_i/\cos\theta_i$ for TE$_r$ and $\eta_{i,\text{eff}}=\eta_i\cos\theta_i$ for TM$_r$ ($i=1,2$). Finally, the surface polarization densities~\eqref{Eq:PM} may also be written in terms of the impedances of the surrounding media as
\begin{subequations}\label{Eq:PM_media}
\begin{align}
 \ve{P}&=\epsilon_0\bar{\bar{\chi}}_\text{ee}(\ve{E}^++\ve{E}^-)/2+\sqrt{\mu_0\epsilon_0}\bar{\bar{\chi}}_\text{em}(\ve{E}^+/\eta_{2,\text{eff}}+\ve{E}^-/\eta_{1,\text{eff}})/2,\\
 \ve{M}&=\sqrt{\epsilon_0/\mu_0}\bar{\bar{\chi}}_\text{me}(\ve{E}^++\ve{E}^-)/2+\bar{\bar{\chi}}_\text{mm}(\ve{E}^+/\eta_{2,\text{eff}}+\ve{E}^-/\eta_{1,\text{eff}})/2.
\end{align}
\end{subequations}
So, the information on the media surrounding the metasurface is not explicitly apparent in the usual form~\eqref{Eq:app_GSTC_two_media} of the GSTCs, but explicitly appears in the specified input ($+$: incident and reflected) and output ($-$: transmitted) fields, which must obviously be specified consistently with Maxwell equations.

\subsection{Derivation of Volume Equivalent Susceptibility}\label{sec:sur_vol_chi}

No softwave is currently available to simulate curved metasurfaces with zero thickness and hence, particularly, zero-thickness spherical metasurfaces. Therefore, we present here a technique allowing to model spherical surface susceptibilities by volume-diluted susceptibilities in a deeply subwavelength spherical shell in order to validate the synthesis presented in Sec.~\ref{sec:synthesis}.

For simplicity, and without loss of generality, consider the case of an isotropic metasurface, with electric susceptibility $\chi_\text{ee}$. In this case, Maxwell-Amp\`{e}re equation reads
\begin{equation}\label{Eq:Maxwell_s_v}
  \nabla \times \ve{H}=j\omega\epsilon_0(1+\chi_\text{ee})\ve{E}.
\end{equation}

\begin{figure}[!h]
    \centering
         \includegraphics[width=1\linewidth, trim={-0.1in 0in -0.4in 0in}]{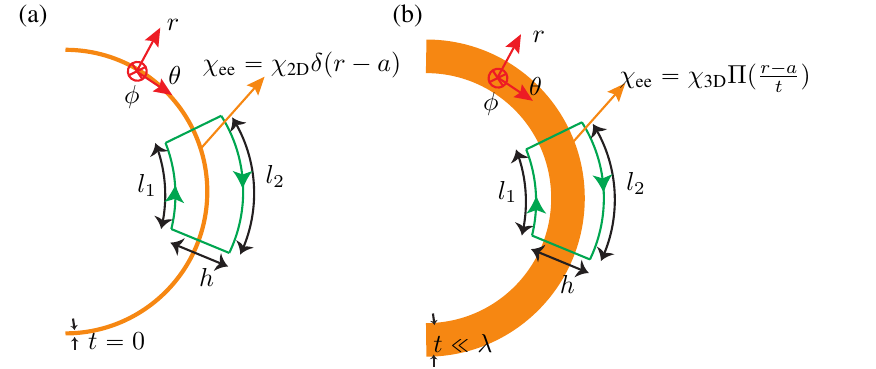}{
        }
        \caption{Integration parameters to derive the equivalence between surface and volume susceptibilities for a spherical metasurface: Eq.~\eqref{Eq:chi_sur_vol}. (a)~Ideal metasurface sheet with zero thickness ($t=0$). (b)~Corresponding metasurface shell with sub-wavelegnth thickness $t\ll\lambda$.}
   \label{FIG:dilution}
\end{figure}

The sought after modeling can be found by integrating this equation for both a metasurface sheet and a subwavelength shell, as shown in \figref{FIG:dilution}, and equating the results.

In the case of the metasurface \emph{sheet} [\figref{FIG:dilution}(a)], we have $\chi_\text{ee}=\chi_\text{2D}\delta(r-a)$, and Eq.~\eqref{Eq:Maxwell_s_v} integrates to

\begin{equation}\label{Eq:Maxwellintegration1}
\oint \ve{H}\cdot d\ve{l}=j\omega\epsilon_0\iint [1+\chi_\text{2D}\delta (r-a)]\ve{E}\cdot d\ve{S},
\end{equation}
which yields
\begin{equation}\label{Eq:surface}
(H_\theta^+l_2-H_\theta^-l_1)=j\omega\epsilon_0(h+\chi_\text{2D})E_\phi\frac{l_2+l_1}{2},
\end{equation}
where the elevation distance has been taken as the average of the elevation distances on both sides of the metasurface ($l_1$ and $l_2$).

In the case of the metasurface \emph{shell} [\figref{FIG:dilution}(b)], we have $\chi_\text{ee}=\chi_\text{3D}\Pi[(r-a)/t]$, where $\Pi(\cdot)$ is the rectangular pulse function, and Eq.~\eqref{Eq:Maxwell_s_v} integrates to
\begin{equation}\label{Eq:Maxwellintegration2}
\oint\ve{H}\cdot\ve{l}=j\omega\epsilon_0\iint\left\{1+\chi_\text{3D}\Pi[(r-a)/t]\right\}\ve{E}\cdot d\ve{S},
\end{equation}
which yields
\begin{equation}\label{Eq:volume}
  (H_\theta^+l_2-H_\theta^-l_1)=j\omega\epsilon_0(h+\chi_\text{3D}t)E_\phi\frac{l_2+l_1}{2}.
\end{equation}

Equating~\eqref{Eq:surface} and~\eqref{Eq:volume}, provides the surface-equivalent volume susceptibility
\begin{equation}\label{Eq:chi_sur_vol}
\chi_\text{3D}=\chi_\text{2D}/t,
\end{equation}
corresponding to the permittivity $\epsilon=1+\chi_\text{ee}/t$, and then also permeability $\mu=1+\chi_\text{mm}/t$, which may be straightforwardly generalized to the anisotropic case in COMSOL.

\end{document}